\documentclass[twoside,draft,reqno,12pt]{amsart}
\usepackage[english]{babel}
\usepackage{amsmath}
\usepackage{amssymb,ulem}
\usepackage{enumerate}

\usepackage{amsfonts}
\usepackage{graphicx}

\usepackage[ citecolor=black, linkcolor=black,
colorlinks, hypertexnames=false, plainpages=false]{hyperref}

\usepackage{xcolor}

\newtheorem{lemma}     {Lemma}[section]
\newtheorem{theorem}   [lemma]{Theorem}
\newtheorem{teorema1}   [lemma]{Theorem}

\newtheorem{cong1}      [lemma]{Conjecture}
\newtheorem{remark}    [lemma]{Remark}
\newtheorem{definition}      [lemma]{Definition}

\numberwithin{equation}{section}

\newcommand{\R}{\mathbb R}

\newcommand{\dis}{\displaystyle}

\newcommand{\mmmintone}[1]{{\dis{\int\kern -.38cm
-}}_{\kern-.21cm\substack{#1}}\;\;}
\newcommand{\mmmintwo}[2]{{\dis{\int\kern -.43cm
-}}_{\kern-.21cm\substack{#1}}^{\substack{#2}}\;\;}
\newcommand{\submint}{{\scriptstyle{\int\kern -.66em -}}}
\newcommand{\submintone}[1]{{\scriptstyle{\int\kern -.66em
-}}_{\scriptscriptstyle{\kern-.21em\substack{#1}}}}
\newcommand{\fracmint}{{\textstyle{\int\kern -.88em -}}}
\newcommand{\fracmintone}[1]{{\textstyle{\int\kern -.88em
-}}_{\scriptscriptstyle{\kern-.21em\substack{#1}}}\;}

\newcommand{\ga}{\gamma}

\newcommand{\La}{\Lambda}

\newcommand{\nada}[1]{}

\def\be{\begin{equation}}
\def\ee{\end{equation}}

\def\supp{{\rm supp}\,}

\def\La{\Lambda}

\def\GG{\mathcal{G}}
\def\CC{\mathcal{C}}
\def\BB{\mathcal{B}}
\def\TT{\mathcal{T}}

\begin{document}
\today

\vskip.5cm
\title[Correlation functions and the Ornstein-Zernike equation]
{Convergence of density expansions of correlation functions and the Ornstein-Zernike equation}

\author{Tobias Kuna}
\address{Tobias Kuna,
Department of Mathematics and Statistics,
University of Reading,
\indent Reading RG6 6AX, UK}
\email{t.kuna@reading.ac.uk}

\author{Dimitrios Tsagkarogiannis}
\address{Dimitrios Tsagkarogiannis,
Department of Mathematics, University of Sussex,
\indent Brighton BN1 9QH, UK}
\email{D.Tsagkarogiannis@sussex.ac.uk}

\begin{abstract}
We prove convergence of the multi-body correlation function as a power series in the density.
We work in the context of the cluster expansion in the canonical ensemble and we obtain
bounds uniform in the volume and the number of particles.
In the thermodynamic limit,
the coefficients are characterized by sums over some class of two-connected graphs.
We introduce the ``direct correlation function'' in the canonical ensemble 
and we prove that in the thermodynamic limit it is given by a convergent power series in the density
with coefficients given by sums over some other class of two-connected graphs.
Furthermore, it satisfies the Ornstein-Zernike equation from which quantified approximations
can be derived.
\end{abstract}


\maketitle


\section{Introduction}
\label{intro}

Correlation functions of interacting particle systems provide
important information of the macroscopic as well as the microscopic properties 
of the system.
In particular, correlation functions of low degree are directly measurable in experiments; e.g. the radial distribution function is the main object of interest in the study of liquids and heterogenous materials.
This was  well captured already in the literature in the 30's, see \cite{K39}  and the references therein.
Around the same period, with the development of power series expansions
by Mayer and his collaborators, \cite{MM},
a direct perturbative representation of correlation functions in terms of integrals over configurations associated to a graphical expansion has been suggested in \cite{MMo}, where the density expansion
of the $n$-body correlation function has been derived.
However, being perturbative expansions around the ideal gas,
 the density expansions of the correlation functions are not expected to 
be valid at the densities of the liquid regime. 
So, it is desirable to {\it ``develop a theory
of classical fluids without using the density expansion formulas"}, \cite{MH}.
Several suggestions were made and we refer to the description given in \cite{MH60} and the references therein.

A candidate for deriving such relations is the
original Ornstein-Zernike (OZ) equation, \cite{OZ}, which, however, 
cannot be solved as an equation as it contains two unknown quantities, namely 
the correlation function and the direct correlation function.  Hence one has to postulate a relation between them, that is a closure scheme. 
A lot of effort has been made in this direction and various
suggestions have appeared.
In \cite{SPY}, G.~Stell relates  the 
most popular closure schemes (such as the Born-Green-Yvon (BGY) hierarchy, \cite{BG, Y}, the Hyper-Netted Chain (HNC) and the Percus-Yevick (PY) equation \cite{PY}) to  graphical expansions and tries to motivate them in this way. However, it is also acknowledged that {\it ``the manipulations involved in obtaining these
infinite sums ... have been carried out in a purely formal way and we have not examined 
the important but difficult questions of convergence and the legitimacy of the
rearrangement of terms"}. 

In the present paper we work on this program by establishing
the convergence of the expansion of the $n$-body correlation function as well as
for the direct correlation function in terms of the density. 
Before we discuss the aforementioned problems and previous works in more detail, let us state the main achievements of this paper:
\begin{enumerate}
\item\label{item1} Absolute convergence of the density expansion for the truncated correlation functions uniformly in the arguments of the correlation function, with essentially the same radius of convergence as for the activity expansion.
\item\label{item2} The graphical representation of the density expansion of the truncated correlation function by some class of two-connected graphs.
\item\label{item4} The convergence of the density expansion (in the thermodynamic limit) of the direct correlation function in the $L^1$ sense in the arguments. Furthermore, we show that this type of convergence implies that the direct correlation function defined via its expansion solves the  OZ equation in the thermodynamic limit. 
\item\label{item5} The order of the error term in  the closure which gives rise to the Percus-Yevick equation is rigorously derived.
\end{enumerate}
Note that the convergence in (\ref{item1}) holds also in $L^1$-sense and one can show that following the line of proof presented in the paper.

The first mathematically rigorous construction of the correlation functions in the thermodynamic limit were obtained in the grand canonical ensemble, \cite{BH}, in the 40's.
Further progress has been made in the 60's with the proofs of the convergence
of the related cluster expansion of the pressure and the virial expansions of the free energy by Penrose, Lebowitz, Groeneveld and Ruelle based on the so-called tree-graph estimate.
In the 50's, also an alternative method appeared, namely the Kirkwood-Salsburg equations for the
grand canonical correlation functions, see \cite{KS, Hill}. 
For a comparison of the above integral equation approaches, 
see \cite{TS} and the references therein.
For the rigorous validity of the solutions of these equations
within the same convergence radius as for the cluster expansions and the convergence of the correlation functions in the context of the grand canonical ensemble
we refer to Ruelle \cite{Rcor}. 
In the canonical ensemble, we refer to 
\cite{BPH}.
After the 70's, the technique of 
cluster expansion has been further developed
and their validity has been established for a large class of different  systems
with the introduction of the abstract polymer model \cite{GK, KP}.
However, for the classical particle model the radius of convergence has not been significantly improved.
We refer to \cite{FP} for a review of the different sufficient conditions for convergence.

For the case of the classical gas, almost all results are based on the grand canonical ensemble as
the techniques that have been used seem to require that one has to
exploit the infinite sum over the number of particles. 
 In order to obtain an expansions in the density, two further steps are required: 
first, some ``inversion" theorem from analytic function theory and second  
a combinatorial relation between graphs, e.g. a ``topological reduction'' in the language of Stell. In exactly this way, in \cite{LP} the virial expansion from the activity expansion was derived using Lagrange inversion and Mayer's combinatorial identities \cite{MM}.
See also \cite{UF}
as well as \cite{JTTU} for the multi-species case.
In \cite{L} this relation between graphs is put in the
systematic context of operations between combinatorial ``species''.
In \cite{LP} it was pointed out that one can derive the convergence of the correlation functions point-wise in the argument as an expansion in the density via the grand canonical cluster expansions. However, in order to work with the expansion, e.g. in order to show that it satisfies the Ornstein-Zernike equation, 
one needs that the series is absolutely convergent with respect to the $L^1$-norm or in the uniform norm in the arguments of the correlation function. This may be possible via an indirect proof using variational inversion formulas
in the spirit of \cite{LP}, but to the best of our knowledge it has not been presented in detail.

In this paper, we follow a direct and natural approach to obtain the density expansion from the canonical ensemble.
In \cite{PT} the validity of the cluster expansion 
in the canonical ensemble has been established
for the free energy
combining the techniques of abstract polymer model and tree-graph estimates for particle systems. Because of the latter, no significant improvement for the radius of convergence
for the virial expansion was achieved. 
Following this development, in this paper we prove the convergence of the
expansions for both the correlation and the direct correlation function by working directly 
in the canonical ensemble. 
Similarly, we prove the validity of the density expansion for the direct correlation
function and show that in the thermodynamic limit the truncated and the direct correlation functions are satisfying the Ornstein-Zernike equation. The latter was also outlined in \cite{AK} for finite range potentials
by examining analyticity in the Fourier space. However, the point-wise convergence established in \cite{LP} is not sufficient to connect the Fourier transform of the series in density with the Fourier transform of its summands.
Our direct approach, apart from being applicable to a larger class of potentials, allows a direct control of the sense of convergence of the expansion with respect to its arguments. 
For example, the validity of the Ornstein-Zernike equation requires 
the control in an integral norm
as in the bound 
\eqref{boundforCbullet}. It turns out that it was essential for the proof that one works with an integral norm allowing to combine translation invariance
and combinatorial cancellations (see Lemma~\ref{lem2.2}).
The main benefit of the direct approach 
is that it elucidates how the derivation of convergence
is related to
the underlying combinatorial
structure of the graphs in the formal computations in Stell.


In liquid state theory,
starting from the Ornstein-Zernike equation, several closures have been suggested. 
Note that these
give rise to approximations that are not as simple as restricting to the leading order, quite the contrary a systematic rule is given on how to select terms from all orders. 
Shortly after, starting from  the grand-canonical ensemble, a
wide range of expansions for different thermodynamic quantities have
been investigated, see the  systematic representation of his and works of others (e.g. \cite{MH60, DD}), by G. Stell in his seminal work in \cite{S}. His approach is mainly based on the tools of functional differentiation
and proper re-summations of the cluster expansion, or ``topological reduction''
as he calls it.
The first was already used in \cite{B}, analogously to  the use of generating functionals
for stochastic processes.
A by-product of the validity of the convergence for the expansions
proved here,
is the proof that these closure schemes have an error of the order of $\rho^2$.
However, going to higher order corrections, as already suggested in \cite{SPY},
it is more complicated. This is left for the future together with the quest of an expansion
that can be valid in the liquid regime.
In this spirit, Hiroike and Morita  suggest that using more complex re-summations 
{\it ``the theory of classical fluids may be constructed with the knowledge of 
the pair distribution function alone, even if a form of the pair interaction
potential is not known."} 
This is also closely related to the inverse or realizability problem, where one
seeks to find a  priori properties of the
correlation function, see \cite{KLS} and the references therein.

The structure of this paper is as follows:
In Section~\ref{sec:2} we present the model and the main results.
Referring to the list given above,
item
\eqref{item1} is based on the definition of the truncated correlation functions via the generating functional for correlation functions, which allows us to 
relate it to the abstract polymer model and
derive the convergence result from the general theorem of cluster expansion cf. Theorem~\ref{thm1} and \ref{thm2}.
This is proved in Section \ref{pf1} and \ref{pf2} respectively.

As expected, the range of convergence is strictly inside the gas phase;
it is the same for both expansions and it can be easily improved along the line of \cite{MP}. 
The proof of item \eqref{item2} is given in Section \ref{pf2}. 
It requires a modification of the cancelations derived in \cite{PT} taking into account the difference in the combinatorial structure. Another crucial property is the  splitting property \eqref{p1.10} which is based on translation invariance. Even though the correlation functions break the translation invariance of the expansion, the splitting property is preserved.
Item \eqref{item4} requires a re-definition of the activity in the abstract polymer representation in order to  show convergence when one of the two arguments of the direct correlation function is considered in the $L^1$ norm, as shown in Section \ref{pf3}. 
The latter modification is essential in order to show the cancelation and the convergence of the expansion in two connected graphs. It is  only in this case that the aforementioned splitting property is preserved.  
In contrast to what one may expect, expansions based on classes of graphs with higher connectivity properties are harder to treat due to the more rigid combinatorial structure. 
We conclude with Sections~\ref{comb}  and \ref{LST}.
In Section~\ref{comb}  we discuss the connections to combinatorial identities. In fact, the different expansions in \cite{S} have a strong combinatorial flavour. 
The results of this paper are applied to liquid state theory in Section~\ref{LST} and are to be investigated further in upcoming  works.  Moreover, as a by-product we also prove item \eqref{item5}.


\section{The model and the results}
\label{sec:2}
We study a system consisting out of $N$ indistinguishable particles described by a configuration 
$ \mathbf{q}:=\{q_1,\ldots,q_N\}$  (where $q_i$ is the position of the $i^{th}$ particle) confined in a box 
$\La(\ell):=(-\frac{\ell}{2},\frac{\ell}{2}]^d
\subset\mathbb{R}^d$ (for some $\ell>0$), which we will also denote for short by $\La$ when we do not
need to explicit the dependence on $\ell$. For simplicity, we consider periodic boundary conditions, that is, we identify opposite sides of the square $\La$ to obtain a torus. The effect of other boundary conditions is left for future studies.
The particles interact via a (translation invariant) pair potential
$V:\R^d\to\R\cup\{\infty\}$, which is stable, integrable at infinity and $V(q)=V(-q)$. A potential $V$ is called stable, whenever
 there exists $B\geq0$ such that:
\be\label{3}
\sum_{1\leq i<j \leq N} V(q_i-q_j) \geq -BN,
\ee
for all $N$ and all $q_1,...,q_N$. 
A potential $V$ is called integrable at infinity, whenever 
\be\label{temper}
C(\beta):= \int_{\mathbb R^d} |e^{-\beta V(q)}-1| dq <\infty .
\ee
The latter condition holds if and only if there exists an $R>0$ such that $ \int_{\mathbb R^d \setminus B_R(0)} | V(q)| dq <\infty$. 
The hard-core potential fulfils all these assumptions and $C(\beta,R) = |B_R(0)|$, the volume of the ball with radius the interaction range $R$.

The energy of the system $H_{\La}$ is defined as 
\be\label{2}
H_{\La}( {\mathbf q})
:=
\sum_{1\leq i<j \leq N} V(q_{i,j}),
\ee
where $q_{i,j}$ denotes among the vectors $q_i -q_j + n \ell$, for $n \in \mathbb{Z}^d$, the one with minimal length. The length of $q_{i,j}$ is equal to the geodesic distance of $q_i$ and $q_j$ on the torus. 

\subsection{Thermodynamic functions and partition functions}

The associated {\it canonical partition function} of the system described above
is given by
\be\label{1a}
Z_{\beta,\La,N}
:=\frac{1}{N!}\int_{\La^N} dq_1\,\ldots dq_N \,e^{-\beta H_{\La}(\mathbf q)}.
\ee 
Given $\rho>0$, the {\it density}, we define the {\it thermodynamic free energy} in the thermodynamic limit by
\be\label{4.1}
f_{\beta}(\rho):=\lim_{\substack{\La\uparrow\mathbb R^{d}, \, N\to\infty, \\ N=\lfloor\rho|\La|\rfloor}} f_{\beta,\La, N}
,\,\,\,{\rm where}
\,\,\,
f_{\beta,\La, N}
:=
-\frac{1}{\beta|\La|}\log Z_{\beta,\La,N},
\ee
where  $|\La|$ is the volume of $\La$. The limit exists for suitable sequences of volumes $\Lambda$ and is actually independent of the boundary condition \cite{FL}.

The associated {\it canonical ensemble} in the volume $\Lambda$ is defined for a measurable set $C \subset \mathbb{R}^{dN}$ by
\be\label{defcanensem}
\mu_{\beta,\La,N}(C)
:=\frac{1}{Z_{\beta,\La,N}}\frac{1}{N!}\int_{\La^N\cap C} dq_1\,\ldots dq_N \,e^{-\beta H_{\La}(\mathbf q)}.
\ee

We introduce some relevant quantities in statistical mechanics to be studied next.
Given a test function $\phi$ we define the Bogoliubov functional $L_{B}(\phi)$ in the canonical ensemble,
in analogy to the definition in the grand-canonical ensemble (by considering the grand-canonical measure restricted to the $N$-particle sector), 
see \cite{B}, equation (2.11): 
\begin{equation}\label{bog}
L_{B}(\phi) :=\int_{\La^N}  \prod_{k=1}^N (1+\phi(q_k))\mu_{\beta,\La,N}(d \mathbf q).
\end{equation}
This is
the generating functional of the correlation functions associated to the canonical ensemble.
In fact, by expanding the product in \eqref{bog} we obtain
\begin{equation}\label{bog1}
L_B(\phi)=\sum_{n=0}^N\frac{1}{n!}\int_{\Lambda^{n}}\phi(q_1)\ldots\phi(q_n)\rho_{\La,N}^{(n)}(q_1,\ldots,q_n)\, dq_1\ldots dq_n,
\end{equation}
where for $n\leq N$ and the points $q_1,\ldots, q_n\in \La$ we have defined the $n${\it-point correlation function
in the canonical ensemble} $\rho_{\La,N}^{(n)}(q_1,\ldots,q_n)$ as:
\begin{equation}\label{twopoint}
\rho^{(n)}_{\La,N}(q_1,\ldots,q_n):=\frac{1}{(N-n)!}\int_{\La^{N-n}} dq_{n+1}\ldots dq_N \frac{1}{Z_{\beta,\La,N}}e^{-\beta H_{\La}(q)}.
\end{equation}
Note that $\rho_{\La,N}^{(0)}=1$ and $\rho^{(1)}_{\La,N}= \frac{N}{|\La|}$. Thus, in the thermodynamic limit 
we obtain $\rho^{(1)}=\rho$.
The existence of the thermodynamic limit $\rho^{(n)}$ for $n \geq 2$, that is the limit 
when $|\La| \uparrow \infty$ with $N=\lfloor\rho|\La|\rfloor$, is more subtle than for thermodynamic quantities like pressure and free energy which are on a logarithmic scale. Analogous results in the grand-canonical ensemble are well-established \cite{R, Rcor}. Furthermore, for small values of the activity, the correlation functions 
can be represented as power series in the activity. 
A by-product  of our analysis below is that we also establish the convergence of the thermodynamic limit in the high-temperature-low-density regime in the canonical ensemble. 
The only related previous result we are aware of is \cite{BPH}.

The logarithm of the Bogoliubov function 
\begin{equation}\label{u}
\log L_B(\phi)=:\sum_{n \geq 1}\frac{1}{n!}\int_{\Lambda^{n}}\phi(q_1)\ldots\phi(q_n)
u^{(n)}_{\La,N}(q_1,\ldots,q_n)\, dq_1\ldots dq_n,
\end{equation}
is the generating function for $u^{(n)}_{\La,N}(q_1,\ldots,q_n)$, the sequence of {\it truncated correlation functions} or {\it Ursell functions}. 
Relation \eqref{u} can be understood as the definition of $u^{(n)}_{\La,N}(q_1,\ldots,q_n)$.

These are the  analogues of the 
cumulants for the sequence of correlation functions. The correlation functions and the {\it Ursell functions} can be related directly via a combinatorial formula by comparing \eqref{bog1} and \eqref{u} and give rise to the usual definition of the Ursell functions, see e.g. \cite{R}, p.87 or \cite{S}, equation ($2$-$8$):
\begin{equation}\label{¡rhoandu}
\rho^{(n)}_{\La,N}(q_1,\ldots,q_n)=\sum_{\{P_1,\ldots,P_k\}\in\Pi(1,\ldots,n)}\prod_{i=1}^k u^{(|P_i|)}_{\La,N}(\underline q_{P_i}),
\end{equation}
where $\Pi(1,\ldots,n)$ is the set of all partitions of $\{1,\ldots,n\}$. For $P_i = \{ j_1, \ldots , j_{|P_i|}\}$, we use the shortcut notation: $\underline q_{P_i} = (q_{j_1}, \ldots , q_{j_{|P_i|}})$.
For example, for $n=2$ we have:
\begin{equation*}
u^{(2)}_{\La,N}(q_1,q_2)=\rho^{(2)}_{\La,N}(q_1,q_2)-\rho_{\La,N}^{(1)}(q_1)\rho_{\La,N}^{(1)}(q_2).
\end{equation*}
We will see that in the thermodynamic limit the functions of $\rho$, $\rho^{(n)}$ and $u^{(n)}$ (the limits of
$\rho^{(n)}_{\La,N}$ and $u^{(n)}_{\La,N}$) have as leading order $\rho^n$. 
Hence, it is common to introduce the following order one functions:
\begin{equation}\label{g}
g^{(n)}_{\La,N}(q_1,\ldots,q_n):=\frac{\rho_{\La,N}^{(n)}(q_1,\ldots,q_n)}{\rho^n}
\end{equation}
and
\begin{equation}\label{h}
h^{(n)}_{\La,N}(q_1,\ldots,q_n):=\frac{u^{(n)}_{\La,N}(q_1,\ldots,q_n)}{\rho^n}.
\end{equation}
Due to the periodic boundary conditions all correlation functions introduced above will be invariant under translation. 
Furthermore, as bounds will be uniform in $\La$ and $N$, it follows that all relations to be described in this subsection will still hold true in the thermodynamic limit.

Next, we concentrate on the case $n=2$. We express all correlation functions as functions of the difference of coordinates $\rho^{(2)} _{\La,N}(q_1-q_2)$, $u^{(2)}_{\La,N}(q_1-q_2)$,
$g^{(2)}_{\La,N}(q_1-q_2)$. 
The latter is known as the {\it radial distribution function} (in case that the potential $V$ is also radially 
symmetric) and $h^{(2)}_{\La,N}(q_1-q_2)$ as the {\it structure function}. Then the following relation holds
\begin{equation}
h^{(2)}_{\La,N}(q_1-q_2)=g^{(2)}_{\La,N}(q_1-q_2)-\left( \frac{N}{\rho |\Lambda|} \right)^2,
\end{equation}
which in the thermodynamic limit simplifies to
\begin{equation}\label{gandh}
h^{(2)}(q_1-q_2)=g^{(2)}(q_1-q_2)-1 .
\end{equation}

Another type of correlation function playing a central role in the theory of liquids is the {\it Ornstein - Zernike direct correlation function} $c(q_{1},q_{2})$. In the thermodynamic limit it is defined via the following relation, 
usually called in the literature as
{\it Ornstein-Zernike equation}:
\begin{equation}\label{OZv1}
h^{(2)}(q_1, q_2)=c(q_1,q_2)+ \int_{\R^d} c(q_1,q_3) h^{(2)}(q_3,q_2)\rho^{(1)}(q_3)\, dq_3.
\end{equation}
The direct correlation function is the building block of the classical theory of fluids, see e.g. \cite{SPY} and the references therein. 
For the case of dilute classical systems with finite-range interactions, it has been investigated in \cite{AK} that
the direct correlation function expressed in terms of its graphical expansion satisfies the Ornstein-Zernike equation by expressing it in the Fourier space (whenever one can interchange the Fourier transform with the series in the density).
The analogous results have been  proved
for the finite range Ising model above the critical temperature  \cite{CIV},
for the Potts model \cite{CIV2},
as well as in the context of 
point processes and the random connection model of percolation \cite{LZ}.
Here, in Theorem~\ref{thm3}, 
working directly in the canonical ensemble and expressing the involved quantities
as graphical expansions,
we prove that the direct correlation function is an absolutely convergent series
in powers of the density and satisfies \eqref{OZv1} in the thermodynamic limit.

In contrast to the grand-canonical ensemble where the activity appears as a parameter in the definition, in the canonical ensemble the density only enters as a parameter in the thermodynamic limit. In finite volume, the right approximation to the density is given by the following (or similar) expression for $n \leq N$:
\begin{equation}\label{P}
P_{N,|\La|}(n):=
\frac{N(N-1)\cdots (N-n+1)}{|\La|^{n}}, \qquad \mbox{for }n\leq N,
\end{equation}
which tends to $\rho^n$ in the thermodynamic limit. For $n >N$ we put $P_{N,|\La|}(n)=0$. For the 
convenience of the reader, the first result is stated without any reference to the polymer expansion to be
considered next. It expresses the correlation functions in terms of these approximated powers of the density and establishes that it convergences to a power series expansion in $\rho$ in the thermodynamic limit.

\begin{theorem}\label{thm1}
There exists a constant 
$c_0:= c_0(\beta,B)>0$, independent of $N$ and $\La$ such that if $\rho\,C(\beta)<c_0$
(with $N=\lfloor \rho |\Lambda | \rfloor$ and $C(\beta)$ as in \eqref{temper}), for a test function $\phi$ we obtain:
\begin{equation}\label{un}
\int_{\Lambda^n} \phi(q_1) \ldots \phi(q_n) u_{\Lambda,N}^{(n)}(q_1,\ldots,q_n)
dq_1\ldots dq_n=\sum_{k\geq 0}
F_{\beta,\La, N}(n,k),
\end{equation}
where 
\begin{equation}\label{F}
F_{\beta,\La, N}(n,k)=
\sum_{m=1}^n
P_{N,|\Lambda|}(m+k)
B_{\beta,\Lambda}(n,m,k).
\end{equation}
The factor
$P_{N,|\Lambda|}(m+k)$ is defined in \eqref{P}, while $B_{\beta,\Lambda}(n,m,k)$ will be given later
in \eqref{B} after introducing the abstract polymer model.
Furthermore, there exist constants $C,c>0$ such that, for every $N$ and $\Lambda$,
the coefficients $F_{\beta,\La, N}(n,k)$, $n\geq 1$, satisfy
\begin{equation}\label{bound}
|F_{\beta,\La, N}(n,k)|\leq C e^{-ck}.
\end{equation}
For $\Lambda \uparrow \mathbb{R}^d$ with $N=\lfloor \rho |\Lambda | \rfloor$ the coefficient $B_{\beta,\Lambda}(n,m,k)$ converges to a limit $\bar{B}_{\beta}(n,k)$ which is determined in \eqref{tryth} and 
the series
\begin{equation}\label{uninfty}
\int_{\mathbb{R}^{dn}} \phi(q_1) \ldots \phi(q_n) u^{(n)}(q_1,\ldots,q_n)
dq_1\ldots dq_n=\rho^{n}\sum_{k\geq 0} \rho^k \bar{B}_{\beta}(n,k)
\end{equation}
is absolutely convergent.
\end{theorem}

\begin{remark}
To prove Theorem~\ref{thm1} we follow the strategy presented in \cite{PT}.
As a result, the radius of convergence or the value of $c_{0}$ is the same as in \cite{PT}.
However, one can easily obtain slightly better values by following the machinery developed in \cite{FP}
and applied in the case of the canonical ensemble in \cite{MP}.
\end{remark}

For convenience we will work with $h_{\Lambda,N}^{(n)}$ (which asymptoticaly coincides with $u_{\Lambda,N}^{(n)}$ up to a power of $\rho$).
The next  step is to identify in $h_{\Lambda,N}^{(n)}$  the
leading order terms that survive in the thermodynamic limit
and show that it converges to a function $h^{(n)}$ 
which is analytic in $\rho$. Furthermore, the limit is uniform in $q_1, \ldots, q_n$.  Up to translation invariance the limit holds also in $L^1$. 
In order to obtain an explicit description of the limiting $h^{(n)}$, we need an explicit asymptotic expression for $B_{\beta,\Lambda}(n,m,k)$ in terms of a graphical representation.
The resulting expression for $h^{(n)}$ was already suggested in \cite{MMo}, \cite{MH60} and \cite{S}, but the question of
convergence of the power series remained open since then and we address it here.
First, we introduce some concepts from combinatorics and graph theory. 
We also denote by $f_{i,j}  := e^{-\beta V(q_i -q_j)}-1$ {\it Mayer's $f$-function. } 
Partially following \cite{L} we define:

\begin{definition}\label{basics}
A \emph{(simple) graph} is a pair $g:= (V(g),E(g))$, where $V(g)$
is the set of \emph {vertices} and  $E(g)$ is the set of  \emph{edges}, with $E(g) \subset \{U\subset V(g) : |U|=2 \}$, $|\cdot|$ denoting the \emph{cardinality} of a set.
A graph $g=(V(g),E(g))$ is said to be \emph{connected}, if for every pair $A,B\subset V(g)$
such that $A\cup B= V(g)$ and 
$A\cap B=\varnothing$, there is an edge $e\in E(g)$ such that $e\cap A\neq\varnothing$
and $e\cap B\neq\varnothing$.
Singletons are considered to be connected.
We use $\CC_V$ to denote the set of connected graphs on the set of vertices 
$V\subset [N]$, where we use the notation $[N]:=\{1,...,N\}$.
\end{definition}

\begin{definition}\label{twoconn}
A \emph{cutpoint} of a connected graph g is a vertex of $g$ whose removal (with the attached edges) yields a disconnected graph. A connected graph is called \emph{$2$-connected} if it has no cutpoint. A \emph{block} in a simple graph is a maximal $2$-connected subgraph. The block-graph of a graph $g$ is a new graph whose vertices are the blocks of $g$ and whose edges correspond to a pair of blocks having a common cutpoint. \end{definition}

Cutpoints are frequently also called \emph{articulation points}.
In this article, we reserve the latter notion for the following slightly more general concept.
We use this terminology in order to stay close to Stell's seminal presentation \cite{S} of these
graphical constructions.

\begin{definition}\label{nodaletc}
Let $k \in \mathbb{N}$, $n \in \mathbb{N}_0$. We consider graphs with $n+k$ vertices, of which the first $n$ vertices are singled out and for simplicity we call them ``white". All other vertices are considered to be ``black".
The set of all such graphs is denoted by $\GG_{n,n+k}$. Single vertices are not considered as graphs. 
Similarly, we denote by $\mathcal C_{n,n+k}$ 
the set of all connected graphs on $n+k$ vertices with $n$ white vertices.

A vertex is called \emph{articulation} vertex if upon its removal the component of which it is part separates into two
or more connected pieces in such a way that at least one piece contains no white vertices.

We denote by $\mathcal B^{\text{AF}}_{n,n+k}$ the subset of $\GG_{n,n+k}$ free of articulation vertices.
\end{definition}

The easiest example to distinguish cutpoint from articulation point is the graph: $1$ (white) - $2$ (black) - $3$ (white),
which is an articulation free graph, but it is not a $2$-connected one, as the vertex $2$
is a cutpoint (but not an articulation point).

This concept of articulation vertices free graph is also crucial for the definition of the so-called direct correlation function, see below in \eqref{c2} and \eqref{c2th}.
Motivated by the distinction between an articulation point and a cutpoint, we introduce the
concept of a {\it nodal} point.

\begin{definition}\label{nodal}
A vertex is a \emph{nodal} vertex if there exists two white vertices in its connected
component, which are different from the first vertex, such that all the paths between that pair of chosen white vertices passes through the first vertex.

We denote by $\mathcal B_{n,n+k}$ the set of all connected graphs 
over $n$ white and $k$ black vertices
free of articulation and of nodal vertices. 
The latter coincides with the collection of all two-connected graphs on $n+k$ vertices
with $n$ white vertices.
\end{definition}

The nodal points are exactly the cutpoints of a graph that are not articulation points.
For a graph $g \in \GG_{n,n+k}$ we define the activity
 \begin{equation}\label{act_with_labels}
\tilde\zeta_{\Lambda}(g,\{1,\ldots,n\}):=
\int \prod_{i=1}^{n+k}\frac{dq_{i}}{|\Lambda|}
 \prod_{\{i,j\}\in E(g)} f_{i,j}
\prod_{i\in \{1,\ldots,n\}}\phi(q_i),
 \end{equation}
 as well as its version without the test function $\phi$, but with dependence 
 on a fixed configuration $q_{1},\ldots,q_{n}$:
 \be\label{act_with_points}
\tilde\zeta^{\bullet}_{\Lambda}(g;q_1,\ldots , q_n):=
\int_{\La^k}  \prod_{j=n+1}^{n+k} dq_j
 \prod_{ \{i,j\} \in E(g)} f_{i,j},
\ee 
where $f_{i,j}  := e^{-\beta V(q_i -q_j)}-1$.
If $\phi$ is compactly supported around some point in $\Lambda$, then
$\tilde\zeta_\Lambda$ scales as $|\Lambda|^{-n-k}$ while $\tilde\zeta^{\bullet}_{\La}$  is of order one. Note also that in this paper
we tend to denote with a $^\bullet$ all quantities
that depend on the positions $q_1,\ldots,q_n$.

\begin{theorem}\label{thm2}
Under the assumptions of Theorem~\ref{thm1}, we can write:
\begin{equation*}
B_{\beta,\Lambda}(n,m,k)=\bar B_{\beta,\Lambda}(n,k)\delta_{n,m}+R_{\beta,\Lambda}(n,m,k),
\end{equation*}
where $\bar B_{\beta,\Lambda}(n,k)$ is of  leading order, that is, the leading order is achieved for $m=n$:
\begin{equation}\label{result}
\bar B_{\beta,\Lambda}(n,k)=\frac{|\Lambda|^{n+k}}{n! k!}\sum_{g\in \mathcal B^{\text AF}_{n,n+k}} \tilde\zeta_{\Lambda}(g,\{1,\ldots,n\})
\end{equation}
and
\begin{equation}\label{remainder}
|R_{\beta,\Lambda}(n,m,k)|\leq\frac{C}{|\Lambda|},
\end{equation}
for all $n$, $k$ and uniformly on $\phi$ (the dependence on $\phi$ is through \eqref{act_with_labels}).
Furthermore, the thermodynamic limit exists and it is given by the following absolutely convergent series in
powers of $\rho$, for all $\rho\,C(\beta)<c_0$ (for the same $c_{0}$ as in Theorem~\ref{thm1}):
\begin{equation}\label{Stell_hn}
h^{(n)}(q_{1},\ldots,q_{n}):=\lim_{\substack{\La\uparrow\mathbb R^{d},N\to\infty, \\ N=\lfloor\rho|\La|\rfloor}}
h^{(n)}_{\Lambda, N}(q_{1},\ldots,q_{n})
=\sum_{k\geq 0} \rho^k \frac{1}{n!k!} \sum_{g \in \mathcal B^{\text{AF}}_{n,n+k}}
\tilde\zeta^{\bullet}(g;q_1,\ldots , q_n),
\end{equation}
where 
\begin{equation}\label{act_with_points_th}
\tilde\zeta^{\bullet}(g;q_1,\ldots , q_n):=\lim_{\Lambda\uparrow\mathbb R^{d}}\tilde\zeta^{\bullet}_{\La}(g;q_1,\ldots , q_n)
=\int_{\mathbb R^{dk}}
\prod_{j=n+1}^{n+k} dq_j
 \prod_{ \{i,j\} \in E(g)} f_{i,j}.
\end{equation}
Moreover, at infinite volume, we have the following bound:
\begin{equation}\label{boundforh}
\sup_{q_{1},\ldots,q_{n}\in\Lambda^{n}} \left| h^{(n)}(q_{1},\ldots,q_{n})
\right|\leq C.
\end{equation}

\end{theorem}

\begin{remark}
Equation \eqref{Stell_hn}
is the representation given in
formula 5-5 in \cite{S}, where it is derived from a formal re-summing of the grand canonical ensemble power series representation in the activity. 
Here the formula is derived directly in the canonical ensemble. In  Stell's words,
{\it ``$h^{(n)}(q_1,\ldots, q_n)$ is the sum of all distinct connected simple graphs consisting of white $1$-circles
labeled by $1,2,\ldots,n$, respectively, some or no black $\rho_1$-circles, and at least one $f$-bond, such that
the graphs are free of articulation circles, i.e., are $1$-irreducible''}.
\end{remark}

For the particular case of two fixed white vertices,
recalling the definition of a nodal point and of the set 
$\mathcal B_{2,n+2}$ we define the {\it direct correlation function}
in the canonical ensemble, i.e., for fixed volume $\Lambda$ and number of
particles $N+2$:
\begin{equation}\label{c2}
c^{(2)}_{\Lambda, N+2}(q_1,q_2):=\sum_{k=0}^N\frac{\rho^k}{k!}\sum_{g\in\mathcal B_{2,2+k}}\tilde\zeta^{\bullet}_{\La}(g; q_1,q_2).
\end{equation}
Then we have the following theorem:

\begin{theorem}\label{thm3}
Under the assumptions of Theorem~\ref{thm1} the direct correlation function in \eqref{c2} 
fulfils the Ornstein-Zernike equation (\ref{OZv1}) up to the order $O(1/|\La |$).
In the thermodynamic limit, $c^{(2)}_{\Lambda, N+2}$ converges to 
\begin{equation}\label{c2th}
c^{(2)}(q_1,q_2):=\sum_{k=0}^\infty\frac{\rho^k}{k!}\sum_{g\in\mathcal B_{2,2+k}}\tilde\zeta^{\bullet}(g; q_1,q_2),
\end{equation} 
which is an analytic function in $\rho$, for $\rho\,C(\beta)<c_0$.
Furthermore, the series \eqref{c2th} converges in the following sense:
\begin{equation}\label{boundforCbullet}
\sup_{q_1\in\Lambda}\int_\Lambda dq_2\, 
\frac{\rho^k}{k!}
\left|
\sum_{g\in\mathcal B_{2,2+k}}\tilde\zeta^{\bullet}_{\La}(g; q_1,q_2)
\right|\leq C e^{-c k},
\end{equation}
uniformly in $\Lambda$ and the limit function fulfils the Ornstein-Zernike equation \eqref{OZv1}.
\end{theorem}

\begin{remark}
As a direct consequence of \eqref{boundforCbullet} we have that
\begin{equation}\label{boundsforOZ}
\sup_{q_{1}\in\Lambda}\int_{\Lambda}dq_{2}\, |c_{\Lambda,N}^{(2)}(q_{1},q_{2})|<\infty
\end{equation}
which, together with \eqref{boundforh} (for $n=2$),
proves that the Ornstein-Zernike equation \eqref{OZv1} is well defined.
\end{remark}

\section{Cluster expansion, proof of Theorem \ref{thm1}.}\label{pf1}

Using the relation between the logarithm of the Bogoliubov function and the truncated correlation functions (Ursell functions), cf.  \eqref{u}, we can express the truncated correlation functions as variational derivatives of the logarithm of an extended partition function:
\begin{equation}\label{un0}
\int \phi(q_1) \ldots \phi(q_n) u^{(n)}(q_1,\ldots,q_n)
dq_1\ldots dq_n
=\frac{\partial^n}{\partial \alpha^n}\log Z_{\beta,\Lambda,N}(\alpha\phi)|_{\alpha=0},
\end{equation}
where
\begin{equation}\label{newpartition}
Z_{\beta,\Lambda, N}(\alpha\phi):=\frac{1}{N!}\int\prod_{i=1}^N(1+\alpha\phi(q_i))e^{-\beta H_{\Lambda}(\underline q)}dq_1\ldots dq_N.
\end{equation}
This follows from the fact that
\begin{equation*}
L_B(\alpha\phi)=\frac{Z_{\beta,\Lambda, N}(\alpha\phi)}{ Z_{\beta,\Lambda, N}(0)}, \qquad Z_{\beta,\Lambda, N}(0)
\equiv Z_{\beta,\Lambda, N}.
\end{equation*}

We define the space $\mathcal V^*_N$ whose elements are all ordered pairs $(V,A)$ where $V\subset\{1,\ldots,N\}$
and $A\subset V$.
We say that two elements $(V_1,A_1)$ and $(V_2,A_2)$ are compatible, and denote it by
$(V_1,A_1)\sim(V_2,A_2)$, if and only if $V_1\sim V_2$, where
two sets $V_1,V_2$ are called {\it compatible} (denoted by $V_1\sim V_2$) 
if $V_1\cap V_2=\varnothing$;
otherwise we call them {\it incompatible} ($\nsim$). 

Then we split \eqref{newpartition} as
\begin{equation}\label{idsplit}
Z_{\beta,\Lambda, N}(\alpha\phi)=\frac{|\Lambda|^{N}}{N!}Z^{\text{int}}_{\beta,\Lambda, N}(\alpha\phi)
\end{equation}
and write
\begin{equation}
Z^{\text{int}}_{\beta,\Lambda, N}(\alpha\phi)=
\sum_{\{(V_1,A_1),\ldots, (V_k, A_k)\}_{\sim}}
\prod_{i=1}^k \zeta_{\Lambda}((V_i,A_i)),
\end{equation}
where
\begin{equation}\label{activities}
\zeta_{\Lambda}\left((V,A)\right):=\alpha^{|A|}
\sum_{g\in\mathcal C_V}\tilde\zeta_\Lambda(g,A),\qquad
\tilde\zeta_\Lambda(g,A):=
\int \frac{d\underline q_{V}}{|\Lambda|^{|V|}}
\prod_{\{i,j\}\in E(g)} f_{i,j}
\prod_{i\in A}\phi(q_i),
\end{equation}
with the latter as already defined in \eqref{act_with_labels}, 
and $d\underline q_V$ is a shorthand for the product measure $\prod_{i\in V}dq_i$.
Hence, we are in the framework of the {\it Abstract Polymer Model} 
which
consists of (i) a set of polymers 
$\mathcal V^*_N$, (ii) a binary symmetric relation $\sim$ of compatibility 
between the polymers (i.e., on $\mathcal V^*_N \times \mathcal V^*_N$)
and (iii) a weight function $\zeta_{\Lambda} : \mathcal V^*_N\to \mathbb{C}$.
We also define the
compatibility graph $\mathbb{G}_{\mathcal V^{*}_N}$ to be
the graph with vertex set $\mathcal V^{*}_N$ and with an edge between two polymers $(V_i,A_i)$
and $(V_j,A_j)$ if and only if they are an incompatible pair.
In this framework we have the following formal relation for the logarithm, which will be justified rigorously in 
Theorem~\ref{thmCE} below (see \cite{KP}, \cite{BZ} and \cite{NOZ}):
\be\label{poly1}
\log Z^{\text{int}}_{\beta,\Lambda, N}(\alpha\phi)=
\log \left( \sum_{\{(V_1,A_1),\ldots, (V_k, A_k)\}_{\sim}}
\prod_{i=1}^k \zeta_\Lambda((V_i,A_i)) \right)
=
\sum_{ I\in \mathcal I(\mathcal V^*_N)} c_{ I}\zeta_{\Lambda}^{ I}
,
\ee
where
\be\label{7.2}
c_I=\frac{1}{I!}\sum_{G\subset\GG_I}(-1)^{|E(G)|}.
\ee

The sum in \eqref{poly1} 
is over the set $\mathcal I(\mathcal V^*_N)$ of all multi-indices $I:\mathcal V^*_N \to\{0,1,\ldots\}$.
We use the shortcut
 $\zeta_{\Lambda}^I:=\prod_{(V,A)}\zeta_{\Lambda}((V,A))^{I((V,A))}$,
 but for notational simplicity in stating the main theorem of cluster expansion, we use the
 notation $\gamma:=(V,A)$ for the generic polymer consisting of the ordered pair $(V,A)\in\mathcal V^*_N$.
 Then, defining
 $\supp I:=\{\ga \in \mathcal V^*_N :\,I(\ga)>0\}$, we denote by
 $\GG_I$ the graph with $\sum_{\ga\in\supp I} I(\ga)$ vertices induced from
 the restricted
 $\mathbb{G}_{\mathcal V^*_N}$ in $\supp I$, 
 by replacing each vertex $\ga$ by the complete graph on
 $I(\ga)$ vertices. 
 Furthermore, the sum in \eqref{7.2} is over all connected subgraphs 
 $G$ of $\GG_I$ spanning the whole set of vertices of $\GG_I$
 and $I!:=\prod_{\ga\in\supp I} I(\ga)!$.
Note that if $I$ is such that  $\GG_{I}$ is not connected (i.e., $I$ is not a \emph{cluster}) then
 $c_{I}=0$.

We state the general theorem as a slightly simplified version of \cite{BZ} and \cite{NOZ},
to which we refer for the proof.

\begin{theorem}[Cluster Expansion]\label{thmCE}
Assume that there are two 
non-negative functions $a,c:\mathcal V^*_N\to\R$ such that for every $\ga\in\mathcal V^*_N$,
$|\zeta_{\Lambda}(\ga)|e^{a(\ga)+c(\ga)}\leq\delta$ 
holds, for some $\delta\in (0,1)$.
Moreover, assume that for every polymer $\ga'$
\be\label{7.4}
\sum_{\ga\nsim\ga'}|\zeta_{\Lambda}(\ga)|e^{a(\ga)+c(\ga)}\leq a(\ga').
\ee
Then, for every polymer $\ga'\in\mathcal V^*_N$, we obtain that
\be\label{7.5}
\sum_{I:\,I(\ga')\geq 1} |c_I\zeta_{\Lambda}^I| e^{\sum_{\ga\in\supp I}I(\ga) c(\ga)}\leq |\zeta_{\Lambda}(\ga')| e^{a(\ga')+c(\ga')},
\ee 
where the coefficients $c_I$ are given in \eqref{7.2}.
\end{theorem}

{\it Proof of Theorem~\ref{thm1}:}
From \eqref{un0}, \eqref{newpartition} and by representing the partition function
by the Abstract Polymer Model, we first check the validity of the
convergence condition \eqref{7.4} of Theorem~\ref{thmCE}.
Using again the notation $(V,A)$ (instead of $\gamma$),
in order to bound the activity $\zeta_\Lambda\left((V,A)\right)$
we use the tree-graph inequality (see the original references \cite{P}, \cite{Br}; here we use 
the particular form given in \cite{PU}, Proposition 6.1 (a)):
\be\label{tgi}
\Big|\sum_{g \in \CC_{n}} \prod_{\{j,k\} \in E(g)} f_{j,k}\Big| \leq 
e^{2\beta B n}
\sum_{T \in\TT_n} \prod_{\{j,k\}\in E(T)} |f_{j,k}|,
\ee
where $\TT_n$ and $\CC_{n}$  are respectively the set of trees and connected graphs with $n$ vertices.
We obtain that
$$
\sum_{(V,A):\, V\ni 1}|\zeta_\Lambda\left((V,A)\right)| e^{c|V|} \leq
$$
\begin{eqnarray*}
& \leq &
\sum_{(V,A):\, V\ni 1}\alpha^{|A|}\|\phi\|_\infty^{|A|}
e^{(2\beta B+c)|V|} |\mathcal T_{|V|}|\frac{|\Lambda|}{|\Lambda|^{|V|}}C(\beta)^{|V|-1}\nonumber\\
& \leq &
e^{(2\beta B+c)}
\sum_{n\geq 2}
\binom{N-1}{n-1}\frac{n^{n-2}}{|\Lambda|^{n-1}}
e^{(2\beta B+c)(n-1)}
C(\beta)^{n-1}
\sum_{A:\, |A|\leq n} (\alpha\|\phi\|_\infty)^{|A|}\nonumber\\
& \leq &
(1+\alpha\|\phi\|_\infty)
e^{(2\beta B+c)}
\sum_{n\geq 2}
\left(
(1+\alpha\|\phi\|_\infty)
e^{(2\beta B+c)}
\frac{N}{|\Lambda|}C(\beta)
\right)^{n-1},
\end{eqnarray*}
where we have also bounded the finite volume integrals $\int_{\Lambda}|f_{j,k}|\, dq_{k}$ by $C(\beta)$,
given in \eqref{temper}.
Hence, for any value of $\alpha$,
by bounding $\frac{N}{|\Lambda|}\leq \rho$ and choosing $\rho C(\beta)$ small enough,
the right hand side is finite being a convergent geometric series;
hence \eqref{7.4} holds.
Then, applying Theorem~\ref{thmCE},
the logarithm of the partition function is an absolutely convergent series \eqref{poly1} 
which we analyse next.
Let 
\begin{equation}\label{nandm}
n:=\sum_{(V,A)\in \supp I} |A| I((V,A))
\qquad
\text{and}
\qquad
m:=\big|\cup_{(V,A)\in\supp I}A\big|.
\end{equation}
Note that $m$ is the number of white vertices and $n$ the number of white vertices counted with their multiplicity, that is the number of times a particular vertex appears in different polymers.
Moreover, let $k$ be the number of the remaining vertices, i.e., all vertices which are in the $V$'s, but not in any of the $A$'s, that is, $\cup_{(V,A)\in\supp I}V=[m+k]$.
As $|V|\geq 2$, then, if $A=\varnothing$ for all $A$, we should have that $2\leq k\leq N$.
Otherwise, if $m=1$ then $k\geq 1$, while for our case of $n\geq 2$, we have that $k\geq 0$ as below. Recall that $[m] = \{ 1, \ldots , m\}$.
We have:
\begin{eqnarray}\label{logZ}
\log Z^{\text{int}}_{\beta,\Lambda,N}(\alpha\phi) & = &
\log Z^{\text{int}}_{\beta,\Lambda, N}(0)
+\sum_{n=1}^{N}\sum_{m=1}^n\sum_{k= 0}^{N-m}
\binom{N}{m+k}\binom{m+k}{m} \alpha^n \hspace{-0.9cm}
\sum_
{\substack{I:\, \cup_{(V,A)\in\supp I}A=[m] \\ 
\cup_{(V,A)\in\supp I}V=[m+k]  \\ 
\sum_{(V,A)\in \supp I} |A| I((V,A))=n}
} 
c_I \zeta_\Lambda^I\nonumber\\
& = &
\log Z^{\text{int}}_{\beta,\Lambda, N}(0)+
\sum_{n\geq 1}\sum_{m=1}^n\sum_{k= 0}^{N-m}
\alpha^n
P_{N,|\Lambda|}(m+k)
B_{\beta,\Lambda}(n,m,k),
\end{eqnarray}
where $P_{N,|\Lambda|}$ is given in \eqref{P},
\begin{equation}\label{B}
B_{\beta,\Lambda}(n,m,k):=
\frac{|\Lambda|^{(m+k)}}{m! k!}\hspace{-0.5cm}
\sum_
{\substack{I:\, \cup_{(V,A)\in\supp I}A=[m] \\ 
\cup_{(V,A)\in\supp I}V=[m+k]  \\ 
\sum_{(V,A)\in \supp I} |A| I((V,A))=n}
} 
 c_I \zeta_\Lambda^I
\end{equation}
and $\mathcal \zeta_{\Lambda}$ is given in \eqref{activities}.

Hence, from \eqref{un0}, taking the $n$-th order derivative in \eqref{logZ} and evaluating at $\alpha=0$, 
we obtain another absolutely convergent series  from which
formula \eqref{un} is proved with $F_{\beta,\Lambda,N}$ as in \eqref{F}.
Furthermore, from \eqref{logZ} and \eqref{7.5} the bound \eqref{bound} follows:
\begin{eqnarray}\label{estforF}
|F_{\beta,\La, N}(n,k)| & \leq & e^{-ck}\sum_{m=1}^{n}
\binom{N}{m+k}\binom{m+k}{m}
\sum_{\substack{I:\, \cup_{(V,A)\in\supp I}A=[m] \\ 
\cup_{(V,A)\in\supp I}V=[m+k]  \\ 
\sum_{(V,A)\in \supp I} |A| I((V,A))=n}
} 
|c_I\zeta_\Lambda^I| e^{ck} \nonumber\\
& \leq &
e^{-ck} \sum_{(V',A'):\, A'\ni \{1\}}\sum_{\substack{I:\\  I((V',A'))\geq 1}}|c_I\zeta_\Lambda^I| e^{c\sum_{(V,A)}I((V,A))|V|}
\nonumber\\
& \leq &
e^{-ck} \sum_{(V',A'):\, A'\ni \{1\}} |\zeta_{\Lambda}((V',A'))| e^{(a+c)|V'|}
\leq nCe^{-ck},
\end{eqnarray}
for some constant $C>0$, uniformly in $N, n , k$ and $\Lambda$. 

Concluding, the proof of \eqref{uninfty} is a consequence of the above uniform bounds
complemented with the investigation of
the infinite volume limit of all terms. 
It will be given in the next section and in particular in formula \eqref{tryth}.
\qed

\section{Leading order terms, proof of Theorem~\ref{thm2}}\label{pf2}

Given \eqref{B} we first identify the terms that will survive in the thermodynamic limit.
We claim that in the thermodynamic limit a summand in \eqref{B} is non-zero only if  for all the polymers $(V,A)$ in $\supp I$, only exactly one has $A\neq \varnothing$.
Indeed, polymers  $(V,A)$ with $A\neq \varnothing$ have activities $\zeta_\Lambda((V_i,A_i))$, $i=1,2$, which are of order $O(1/|\Lambda|^{|V_i|})$, whereas, polymers of the type $(V,\varnothing)$
have associated activities $\zeta_\Lambda((V,\varnothing))$ of order $O(\frac{1}{|\Lambda|^{|V|-1}})$. As each polymer has at least one vertex in common with some other polymer, 
this implies the claim by power counting. More precisely, let us consider the following case: suppose the contrary is true and let $(V_1,A_1)$ and $(V_2,A_2)$ be two polymers with both $A_1\neq \varnothing$ and $A_2\neq \varnothing$. 
Moreover, the two polymers are
connected with each other either directly (sharing a label) or via other polymers of the type $(V,\varnothing)$.
If $V_1\cap V_2\neq \varnothing$ then $B_\Lambda$ (given in \eqref{B}) 
is of order $O(\frac{1}{|\Lambda|})$.
The same is true if they are connected via other polymers of the type $(V,\varnothing)$. 
In order to show it, let us assume (without loss of generality)
that there is only one such connecting polymer.
As for the latter the activity is $\zeta_\Lambda((V,\varnothing))=\frac{1}{|\Lambda|^{|V|-1}}$,
we obtain that, again,
the corresponding term in $B_{\beta,\Lambda}$ is of the order of:
\begin{equation*}
|\Lambda|^{|V_1|+|V_2|+|V|-2}  \frac{1}{|\Lambda|^{|V_1|}} 
\frac{1}{|\Lambda|^{|V_2|}} \frac{1}{|\Lambda|^{|V|-1}}=\frac{1}{|\Lambda|}.
\end{equation*}
Hence, the structure of the leading term at the level of the multi-indices is quite simple: 
only one polymer, call it $(V_{0},A_{0})$ has $A_{0}\neq\varnothing$.
Then, for all other polymers with $A=\varnothing$ 
we can have a further structure
as explained below (and as in \cite{PT}).
Since it is always true that the total number of labels ($m+k$) should satisfy
$m+k\leq \sum_{V\in\supp I}(|V|-1)+1$
(due to the fact that each $(V,A)$ should be incompatible with at least one of the other polymers, 
i.e., have at least one common label and $V_{0}\cup \bigcup_{(V,A)\in\supp I, \, V\neq V_{0}}V=[m+k]$), overall we have:
\begin{eqnarray}
&& I((V,A))=1,\,\forall (V,A)\in\supp I,\,\,\, \text{and}\label{p1.3}\\
&& m+k= |V_{0}|+\sum_{(V,A)\in\supp I, \, V\neq V_{0}}(|V|-1)\label{p1.4}.
\end{eqnarray}
Hence, we restrict the summation over multi-indices in this subclass 
satisfying properties \eqref{p1.3}, \eqref{p1.4} and containing
only one polymer $(V_{0},A_{0})$ with $A_{0}\neq\varnothing$. We denote this fact
by adding a superscript  $*$ at the sum as e.g. in \eqref{Bbar} below.
The polymers $(V,\varnothing)$ can be attached to the polymer with $A\neq\varnothing$ either on a vertex not in $A$ (a black circle in the terminology of Stell)
or in a vertex in $A$ (a white circle in the terminology of Stell). 
In order to visualize the last case, we give the following example: consider the
following multi-index $I$: $I$ is equal to one on the two polymers $(\{1,2\},\{1,2\})$ and $(\{1,3\},\varnothing)$, zero otherwise.  The two polymers  intersect in the label $1$.
We have:
\begin{eqnarray*}
|\Lambda|^3\zeta_{\Lambda}^I & = & 
|\Lambda|^3\alpha^2\int \phi(q_1)\phi(q_2)f_{1,2}(q_1-q_2)\frac{dq_1}{|\Lambda|}\frac{dq_2}{|\Lambda|}
\cdot \int f_{1,3}(q_1-q_3)\frac{dq_1}{|\Lambda|}\frac{dq_3}{|\Lambda|}\\
& = &
\alpha^2\int \phi(q_1)\phi(q_2)f_{1,2}(q_1-q_2)dq_1 dq_2
\cdot \frac{1}{|\Lambda|}\int f_{1,3}(q_1-q_3)dq_1 dq_3.
\\
& = &
\alpha^2\int \phi(q_1)\phi(q_2)f_{1,2}(q_1-q_2)dq_1 dq_2
\cdot \int f_{1,3}(q_3) dq_3.
\end{eqnarray*}
As we will explain later, this term will be canceled by one summand from the term $|\Lambda|^3\zeta_\Lambda^{I'}$, with $I'$ being the multi-index which is one only 
on the polymer  $(\{1,2,3\},\{1,2\})$ and in particular with the summand in $|\Lambda|^3\zeta_\Lambda^{I'}$ which is associated with the graph on $\{1,2,3\}$ with exactly two edges  $\{3,1\}$ 
and $\{1,2\}$. Let us start with the formal proof for these cancelations.

\medskip

{\it Proof of Theorem~\ref{thm2}:}
Following the discussion above,
we split $B_{\beta,\Lambda}$ from \eqref{B} as follows:
\begin{equation}\label{split}
B_{\beta,\Lambda}(n,m,k)=\bar B_{\beta,\Lambda}(n,k)\delta_{n,m}+R_{\beta,\Lambda}(n,m,k),
\end{equation}
where
\begin{equation}\label{Bbar}
\bar B_{\beta,\Lambda}(n,k):=
\frac{|\Lambda|^{(n+k)}}{n! k!}
\sum^*_{I:\, A(I)=[n+k]}
 c_I \zeta_\Lambda^I
\end{equation}
and
\begin{equation}\label{AofI}
A(I):=\cup_{V\in\supp I}V.
\end{equation}
Recall that the superscript $*$ indicates that the sum is over all multi-indices that satisfy properties \eqref{p1.3}, \eqref{p1.4} and that contain
only one polymer with $A\neq\varnothing$, for which we have already chosen its
labels and we call it $A_0:=\{1,\ldots,n\}$.
For this reason, we can now consider multi-indices in 
$\mathcal I(\mathcal V_{n,k})$, where
the class $\mathcal V_{n,k}$ consists of all
subsets of the labels corresponding to 
the white vertices $\{1, \ldots , n\}$ and the black vertices $\{n+1, \ldots , n+k\}$. 
The new polymers either they contain $A_0$ or they intersect it at most one point.
Therefore, in the new set-up with $I\in\mathcal I(\mathcal V_{n,k})$ 
the conditions \eqref{p1.3} and \eqref{p1.4} can be rewritten as
\begin{eqnarray}
&& I(V)=1,\,\forall V\in\supp I,\,\,\, \text{and}\label{p2.3}\\
&& n+k= |V_0|+\sum_{V\in\supp I, \, V\neq V_0}(|V|-1),\label{p2.4}
\end{eqnarray}
where $V_0\supset A_0$
and we still refer to them by a $*$ over the sum.
Moreover, the term $R_{\beta,\Lambda}(n,m,k)$ in \eqref{split} 
consists of lower order terms $1/|\Lambda|$.
The proof that also their sum is of order $1/|\Lambda|$ is a bit more delicate, but it has been done
in \cite{PT2}, Lemma 6.1 and 6.2 to which we refer for the details.

The next step is to investigate cancellations that take place in finite volume.
These originate from the fact that in the sum in \eqref{Bbar}, the activity
function corresponding to a given structure (graph) might appear in several multi-indices
multiplied with different combinatorial coefficients and as a result they may 
cancel with each other exactly.
To implement this step, we fix a graph and we sum over all multi-indices
that can produce it (also compatible with the previous restriction, in particular  that all white vertices
have to be in one polymer)
and apply Corollary~1 in
\cite{PT}.
However, this corollary can only be applied directly for the case of only ``black" vertices.
For example, the graph $1$ (white) - $2$ (black) - $3$ (white)
is not canceled. The vertex $2$, even though it is a cutpoint, it is not an articulation point
as it is linked to only ``white'' vertices.
Indeed, in this case we do not have cancellations as the vertices $1$ and $3$ are white and hence the only 
cluster of polymers in the sum \eqref{Bbar} which contains this graph is $(\{1,2,3\} , \{1,2\}))$ since all whites
have to be in only one polymer.
This is one example of a graph that survives the cancelation.

In the next lemma we give a substantial account of these cancelations and show that \eqref{Bbar} can be expressed as in \cite{S}.

\begin{lemma}\label{Vtog}
For all $n\geq 2, k\geq 1$ and $\Lambda$ large enough, \eqref{Bbar} is equal to
\begin{equation*}
\bar B_{\beta,\Lambda}(n,k)=\frac{|\Lambda|^{n+k}}{n! k!}\sum_{g\in \mathcal B^{\text AF}_{n,n+k}} \tilde\zeta_\Lambda(g,\{1,\ldots,n\}),
\end{equation*}
as it has been announced in \eqref{result}.
\end{lemma}

The proof of Lemma~\ref{Vtog} will be given after concluding the proof of the theorem.
The next challenge is to extract bounds on
the quantity $h^{(n)}_{\Lambda, N}(q_{1},\ldots,q_{n})$.
To this end, we need to interchange the integrals over $q_1,\ldots,q_n$ with the sum over $k$ in the thermodynamic limit, hence
we need to prove convergence of the cluster expansion with activities being functions of $q_1,\ldots,q_n$
in an appropriate norm.
From \eqref{F} using the splitting \eqref{split} we have:
\begin{equation}\label{Fsplit}
F_{\beta,\Lambda,N}(n,k)=P_{N,|\Lambda|}(n+k)\bar B_{\beta,\Lambda}(n,k)
+\sum_{m=1}^{n-1}P_{N,|\Lambda|}(m+k)R_{\beta,\Lambda}(n,m,k),
\end{equation}
where the second term is vanishing in the limit $\Lambda\uparrow\mathbb R^d$.
Substituting in \eqref{un} we obtain:
\begin{eqnarray}\label{try}
&&\int_{\Lambda^n}\prod_{i=1}^n \left(dq_i \, \phi(q_i)\right)
\rho^n h^{(n)}_{\Lambda,N}(q_1,\ldots,q_n)   = \nonumber \\
&& \sum_{k\geq 0}P_{N,|\Lambda|}(n+k)
\frac{1}{n! k!}
\sum_{g\in\mathcal B^{\text{AF}}_{n,n+k}}
\int_{\La^{n+k}}  \prod_{j=1}^{n+k} dq_j
 \prod_{ \{i,j\} \in E(g)} f_{i,j}
 \prod_{i=1}^{n}\phi(q_i)\nonumber\\
&& +\sum_{k\geq 0}\sum_{m=1}^{n-1}P_{N,|\Lambda|}(m+k)R_{\beta,\Lambda}(n,m,k).
\end{eqnarray}
Then, having the bounds \eqref{bound} and \eqref{remainder}
we can take the thermodynamic limit on the right hand side of  \eqref{try} and obtain:
\begin{equation}\label{tryth}
\sum_{k\geq 0}
\rho^n
\rho^k
\bar B_{\beta}(n,k),
\quad
\bar B_{\beta}(n,k):=
\int_{\mathbb R^{dn}}\prod_{i=1}^n \left(dq_i \, \phi(q_i)\right)
\frac{1}{n! k!}
\sum_{g\in\mathcal B^{\text{AF}}_{n,n+k}}
\int_{\mathbb R^{dk}}\prod_{i=1}^k dq_{n+i}
\prod_{ \{i,j\} \in E(g)} f_{i,j},
\end{equation}
which implies \eqref{uninfty} in Theorem~\ref{thm1}
and gives an explicit formula for $\bar B_{\beta}(n,k)$.
In order to obtain \eqref{Stell_hn} 
we need to go one step further and show that we
can exchange the sum over $k$ and the integral over $dq_1\ldots dq_n$.
This is the content of the next lemma where we choose to work
in the finite volume case, since we will need it in the sequel.
\begin{lemma}\label{exchange}
For any $n\geq 2$ and $k\geq 1$ we have that
\begin{equation}\label{boundforBbullet}
P_{N,|\Lambda|}(n+k)
  \frac{1}{n! k!}
 \int_{\Lambda^{k}}  \prod_{j=1}^{k} dq_{n+j}
\left| \sum_{g\in\mathcal B^{\text{AF}}_{n,n+k}}
 \prod_{ \{i,j\} \in E(g)} f_{i,j} \right| \leq C \rho^{n}e^{-ck},
\end{equation}
for some positive constants $c,C$ independent of $k$, $N$ and $\Lambda$, with $N=\lfloor \rho |\Lambda | \rfloor$.

\end{lemma}

The proof of Lemma~\ref{exchange} will be given at the end of this section.
Since the bound \eqref{boundforBbullet} is uniform in the volume $\Lambda$,
we can pass to the limit $\Lambda\uparrow\infty$ and 
prove \eqref{boundforh},
concluding the proof of Theorem~\ref{thm2}. \qed

\bigskip

We conclude this section with the proofs of the two lemmas.

{\it Proof of Lemma~\ref{Vtog}.} We rearrange the finite sum in \eqref{Bbar} by first fixing a graph $g\in\mathcal C_{n,n+k}$
and then summing over all multi-indices in the new space
$\mathcal I(\mathcal V_{n,k})$ that can produce such graph. 
Hence, given $g\in\mathcal C_{n,n+k}$, we identify the articulation points and define the set of graphs 
$\mathbb B(g):=\{b_{0}, b_1,\ldots,b_r\}$ where the $b_i$'s are the components free of articulation vertices.
Notice that one of them, $b_0$ (without loss of generality), contains all white vertices with labels in
$A_0$.
We denote by $\mathcal F_{\nsim}(g)$ the collection 
of all $F\subset \mathbb B(g)$ such that $\cup_{b\in F}b$ is a connected graph,
where we use the notation $\cup_{b\in F}b:=(\cup_{b\in F}V(b),\cup_{b\in F}E(b))$
for the union of graphs.
We also define $\mathcal H(g)$ to be the collection of all such graphs:
\begin{equation}
\mathcal H(g):= \{ g': g' =\bigcup_{b\in F}b,  F \in \mathcal F_{\nsim}(g) \}.
\end{equation}
Similarly,
\begin{equation}\label{may9}
\mathcal A(g):= \{ V(g'),\,g'\in\mathcal H(g) \}
\end{equation}
is the collection of the corresponding subsets of the set of labels.
We use the shortcut $I\sim g$ for the class 
$I:\,\supp I\subset\mathcal A(g)$ with $A(I)=V(g)$,
$|V\cap V'|=1, \forall V, V'\in\supp I$  and each edge of $g$ is contained in 
some polymer $V$ with $I(V)>0$. We have:
\begin{equation}\label{beforetheend}
\bar B_{\beta,\Lambda}(n,k)
=\frac{|\Lambda|^{n+k}}{n! k!}\sum_{g\in \mathcal C_{n,n+k}}
\tilde\zeta_\Lambda(g,\{1,\ldots,n\})
\sum^*_{I\sim g} c_I,
\end{equation}
where we recall that the sum $*$ is over all multi-indices that satisfy \eqref{p2.3}, \eqref{p2.4}
and that contain
only one polymer with $V\supset A_0$, $A_0=\{1,\ldots,n\}$. 
Note that this sum is finite.
Then, in order to obtain \eqref{result}, we show that the sum of multi-indices 
$\sum^*_{I\sim g} c_I$ is one if
$g\in \mathcal B^{\text AF}_{n,n+k}$ 
and zero otherwise,
as it will be proved in \eqref{lmay3}.

To do that, we follow 
the corresponding proof in \cite{PT}. We give here the necessary modifications.
The key property for the cancellations is the fact that for any $g'\in\mathcal H(g)$,
with $g' =\bigcup_{b\in F}b$ for some $F \in \mathcal F_{\nsim}(g)$, the following
{\it factorization} holds 
\begin{equation}\label{p1.10}
\tilde\zeta_{\La}(g')=\prod_{b\in F}\tilde\zeta_{\La}(b),
\end{equation}
for all finite $\La$. 
Note that for simplicity we have used the notation
$\tilde\zeta_{\La}(g):=\tilde\zeta_{\La}(g,A_0)$, if $V(g)\supset A_0$
and $\tilde\zeta_{\La}(g):=\tilde\zeta_{\La}(g,\varnothing)$ otherwise.
The relation \eqref{p1.10} is due to the fact that the intersection points of the articulation vertex free components $b$
in $g'$ are articulation points (for $g'$) and
that for the integration in $\tilde\zeta_{\La}$ we assume periodic boundary conditions.
Moreover, all white vertices are contained in $b_0$. 
Notice that if we had white vertices in different components,
then \eqref{p1.10} would not be true.
Then the main result in \cite{PT}, Lemma 2 still holds true:
\begin{lemma}\label{may}
For any $V^*\in\mathcal V_{n,k}$ and any $g\in\CC_{V^*}$,
let $\mathbb{B}(g)=\{b_{0}, b_1,...,b_k\}$ be the set of its articulation vertex free components.
Thus 
there exists $\ell_0>0$
such that for all $\ell>\ell_0$
the coefficient multiplying the monomials
$\tilde \zeta_{\La}(b_0)^{n_0}, \tilde \zeta_{\La}(b_1)^{n_1}, \ldots 
\tilde\zeta_{\La}(b_k)^{n_k}$
(where $\La\equiv\La(\ell)$), 
for any $n_i\in \{1,2,...\}$, $i=0, 1,\ldots,k$,
in the series $\sum_{I:\,A(I)\subset 
V^*} c_I \zeta_{\La}^I$ with $\zeta_{\La}(V)=\sum_{g'\in\CC_V}\tilde\zeta_{\La}(g')$,
is equal to zero except when $k=0$, i.e., when $g$ is itself an articulation vertex free graph.
\end{lemma}

The only modification in the proof with respect to \cite{PT} is when we check the convergence of the new
cluster expansion, in equation (47). The presence of the white vertices makes it even
easier since we win a power of $|\Lambda|$ because of non translation invariance, therefore
we refrain from repeating the proof here and we refer the reader to \cite{PT}.

Thus, since we know that in \eqref{beforetheend} the component $b_0$
has to appear in each summand and since by Lemma \ref{may} there should be only one
component, then the only non-zero contribution comes from the articulation vertex free component,
i.e., $g\in\mathcal B^{\text{AF}}_{n,n+k}$.
In other words, we have that for every $g\in \mathcal C_{n,n+k}\cap(\mathcal B^{\text{AF}}_{n,n+k})^c$,
\begin{equation}\label{lmay3}
\sum_{\substack{I:\,\supp I\subset\mathcal A(g),\,A(I)=V^* \\
|V\cap V'|=1, \forall V,V'\in\supp I}}c_I=0
\end{equation}
and $=1$, otherwise.
Notice the difference with respect to \cite{PT}:
here, the element $b_0 \in\mathbb B(g)$ as it appears in $\mathcal A(g)$ (via $\mathcal H(g)$, defined above) is special and consists of articulation free graphs in their new definition
within the presence of ``white'' vertices.
This concludes the proof of Lemma~\ref{Vtog}.
\qed

\medskip

{\it Proof of Lemma~\ref{exchange}.} 
Recall the use of the shortcut $I\sim g$ for the multi-indices in
$\mathcal I(\mathcal V_{n,k})$,
as in \eqref{beforetheend}.
Then, we can write the left hand side of \eqref{boundforBbullet} as follows:
\begin{eqnarray}
&& \frac{N(N-1)\ldots (N-(n+k)+1)}{|\Lambda|^{n+k}}\frac{1}{n! k!}
 \int_{\Lambda^{k}}  \prod_{j=1}^{k} dq_{n+j}
\left| \sum_{g\in\mathcal C_{n,n+k}}
 \prod_{ \{i,j\} \in E(g)} f_{i,j} \sum^{*}_{I\sim g} c_I \right| \nonumber \\
 &=&
 \binom{N}{n+k}\binom{n+k}{n}
 \int_{\Lambda^{k}}  \prod_{j=1}^{k} dq_{n+j}
 \frac{1}{|\Lambda|^{n+k}}
\left| \sum^{*}_{\substack{I\in\mathcal I(\mathcal V_{n,k})\\A(I) = [n+k]}} c_I \sum_{\substack{g\in\mathcal C_{n,n+k}: \\ g \sim I}}
 \prod_{ \{i,j\} \in E(g)} f_{i,j} \right| \nonumber \\
&\leq&
\sum^{*}_{\substack{I\in\mathcal I(\mathcal V_{n,k})\\A(I) = [n+k]}} |c_I|  
 \binom{N}{n+k}\binom{n+k}{n}
 \int_{\Lambda^{k}}  \prod_{j=1}^{k} dq_{n+j}
  \frac{1}{|\Lambda|^{n+k}}
\left| \sum_{\substack{g\in\mathcal C_{n,n+k}: \\ g \sim I}}
 \prod_{ \{i,j\} \in E(g)} f_{i,j} \right| , \label{eqclustq}
\end{eqnarray}

where the class $g\sim I$ consists 
of all graphs that can be constructed as follows: for each 
$V \in \supp I $, choose a graph $g_V \in \mathcal C_{V}$ and $g$ is obtained by gluing the graphs $g_V$ and $g_{V'}$ at the unique intersection point $V \cap V'$.
Let $V_0, V_1, \ldots , V_r$ be the polymers in the support of a given $I$, i.e., with $I(V_i)>0$, $i=0,\ldots,r$. 
Without loss of generality we suppose that $V_0$ is the (only) polymer that
contains $A_0:=\{1,\ldots,n\}$, the set of the labels corresponding to the white vertices.
We can write:

\begin{equation*}
\left| \sum_{\substack{g\in\mathcal C_{n,n+k}: \\ g \sim I}}
 \prod_{ \{i,j\} \in E(g)} f_{i,j}\right| =\left| \sum_{g_0\in\mathcal C_{V_0}}\prod_{ \{i,j\} \in E(g_j)} f_{i,j}\right| \ \prod_{j=1}^r \left|\sum_{g_j\in\mathcal C_{V_j}}\prod_{ \{i,j\} \in E(g_j)} f_{i,j}\right| ,
\end{equation*}
as each of the polymers $V_1, \ldots , V_r$ intersects with $V_0$ at most at one label.
Alluding to the constraints \eqref{p2.3} and \eqref{p2.4} 
we split the integral as follows:
\begin{equation}\label{match}
 \int_{\Lambda^{k}} \prod_{j=1}^{k} dq_{n+j} 
  \frac{1}{|\Lambda|^{n+k}}
 \left| \sum_{\substack{g\in\mathcal C_{n,n+k}: 
 \\ g \sim I}}
 \prod_{ \{i,j\} \in E(g)} f_{i,j}\right|
 \leq 
\frac{1}{|\Lambda|^{n}}
 \prod_{j=0}^{r}\hat{\zeta}_\Lambda^\bullet(V),
\end{equation}
where, we have introduced the notation:
\begin{equation*}\label{eqzetabullet}
\hat{\zeta}_\Lambda^\bullet(V) 
:=  
 \left\{ \begin{array}{ll}  
 \int_{\Lambda^{|V\setminus A_0|}} 
\frac{d\underline q_{V\setminus A_0}}{|\Lambda|^{|V\setminus A_0|}}\left|\sum_{g\in\mathcal C_{V}}\prod_{ \{i,j\} \in E(g)} f_{i,j}\right|   ,
  & \mbox{if } V \supset A_{0} ,\\
 \int_{\Lambda^{|V|}} \prod_{j\in V} \frac{dq_j}{|\Lambda|} \left|\sum_{g\in\mathcal C_{V}}\prod_{ \{i,j\} \in E(g)} f_{i,j}\right|  ,
  & \mbox{if } |V \cap A_{0}| \in \{ 0, 1\},\\
  0, & \mbox{otherwise.}
  \end{array} \right.
\end{equation*}
Note that these activities differ from the ones in \eqref{activities}
by not having the test functions $\phi$ inside of the integral, but instead  some
fixed configurations $\underline q_{A_0}$ (which we indicate by the $^\bullet$). 
Thus, we can bound \eqref{eqclustq} using \eqref{match} and the definition of the binomial coefficient:
\begin{equation}\label{froml1toNfirst}
\frac{\rho^n}{n!}
\sum_{\substack{I\in\mathcal I(\mathcal V_{n,k})\\A(I) = [n+k]}}^{*}|c_{I}|\binom{N-n}{k}
\prod_{j=0}^r|\hat\zeta_{\Lambda}^{\bullet}(V)|\leq
\frac{\rho^n}{n!}
\sum_{\substack{A\subset [N-n]\\ |A|=k}}
\sum_{\substack{I: \\ A(I)=A\cup A_0}}
^{*} |c_I|
|\hat\zeta_\Lambda^\bullet|^I.
\end{equation}
Then, it is easy to show that the abstract polymer model in $\mathcal V_{n,N-n}$
(with $n$ white labels and $N-n$ black)
with compatibility condition $V\sim V'$ if and only if $V\cap V'=\varnothing$
and activities $\hat\zeta_\Lambda^\bullet$,
satisfies the hypothesis \eqref{7.4} of Theorem~\ref{thmCE}. 
To show it,
for the case $V_{0}\supset A_{0}$
we have:
\begin{equation}\label{temp2}
\sup_{\underline q_{A_{0}} \in \Lambda^{|A_{0}|}} |\hat{\zeta}_\Lambda^\bullet(V_0)|
\leq e^{2\beta B |V_{0}|}\sum_{\tau\in\mathcal T_{V_{0}}} 
\sup_{\underline q_{A_{0}} \in \Lambda^{|A_{0}|}} \int d\underline q_{V_{0}\setminus A_{0}} 
\prod_{ \{i,j\} \in E(\tau)} \left|f_{i,j}\right|.
\end{equation}
Considering one of the labels in $A_{0}$ as the root, we take the supremum of $f_{i,j}$ for any edge which has another  label from $A_{0}$ as a vertex further away from the root. This will give a contribution of 
$\|f_{i,j}\|_{\infty}$ for each such edge.
The  remaining vertices give a contribution $C(\beta)$.
Overall, we bound \eqref{temp2} by
\begin{equation}
\leq e^{2\beta B |V_{0}|} 
|\mathcal{T}_{|V_{0}|}| \left( \|f_{i,j}\|_{\infty} \vee C(\beta)\right)^{|A_{0}|-1} C(\beta)^{|V_0| - |A_0|},
\end{equation}
where $s\vee t$ denotes the maximum of the two numbers $s,t$.
Similarly, for the case $|V\cap A_{0}|\in\{0,1\}$, we have:
\begin{equation}\label{boundforzeta}
|\hat{\zeta}_\Lambda^\bullet(V)|\leq
|\mathcal{T}_{|V|}| (e^{2\beta B} C(\beta))^{|V|}.
\end{equation}
With these bounds it is easy to show that \eqref{7.4} holds.
 Then, Theorem~\ref{thmCE} can be applied obtaining an absolutely
convergent series $\sum_{I\in\mathcal I(\mathcal V_{n,k})}c_I (\hat\zeta_\Lambda^\bullet)^I$,
equal to the logarithm of some abstract polymer model partition function,
but which does not necessarily correspond to some correlation function due to the absolute value in \eqref{eqzetabullet}.
Thus, from \eqref{froml1toNfirst}, using \eqref{7.5},
we obtain that
\be
\sum^*_{\substack{I\in\mathcal I(\mathcal V_{n,N-n}) \\  |A(I)\setminus A_{0}|=k}}
|c_I| |\hat\zeta^\bullet_\Lambda|^I
\leq e^{-c k}
\sum^*_{\substack{I\in\mathcal I(\mathcal V_{n,N-n}) \\  |A(I)\setminus A_{0}|=k}}
|c_I|  |\hat\zeta^\bullet_\Lambda|^I e^{c k} \leq C e^{-c k},
\ee
for some $C>0$ as in \eqref{estforF}, depending on $n$.
\qed

\section{Direct correlation function, proof of Theorem~\ref{thm3}}\label{pf3}

Using \eqref{un}, Theorem~\ref{thm2} and definition \eqref{act_with_points} the leading order of the second Ursell function can be expressed as follows:
\begin{eqnarray}\label{u2}
&& \int_{\Lambda^2} dq_1 dq_2 \,\phi(q_1) \phi(q_2)
u^{(2)}_{\La, N}(q_1,q_2)=\nonumber\\
&&\int_{\Lambda^2}dq_1 dq_2 \,
 \phi(q_1) \phi(q_2) \sum_{k\geq 0}
P_{N,|\Lambda|}(2+k)
\frac{1}{2! k!}\sum_{g\in \mathcal B^{\text AF}_{2,2+k}} \tilde\zeta^{\bullet}_\Lambda(g; q_1, q_2)
+O\left(\frac{1}{|\Lambda|}\right).
\end{eqnarray}
In order to derive the Ornstein-Zernike equation in the canonical ensemble, we split the graphs in 
the right hand side of \eqref{u2} at the  nodal points (recall Definition~\ref{nodal}).
These are the points through which pass all paths joining $q_1$ to $q_2$, hence we can order them.
Given $g\in \mathcal B^{\text AF}_{2,2+k}$, we choose the first nodal point starting from $q_1$ and call its label $j$. 
Note that by the definition of articulation points, $j \neq 1,2$.
Upon the removal of this point the graph $g$ splits into two connected components: $g_1$ with $l+2$ vertices and $g_2$ with $k-l+1$ vertices with the only
common vertex being the one with label $j$. Note that $g_1$ contains $q_1$ and $g_2$ contains $q_2$.
Since $q_j$ is the location of a nodal point, we can write 
\begin{equation*}
\tilde\zeta^{\bullet}_\Lambda(g;q_1,q_2)
=\int_\Lambda dq_j \, \tilde\zeta^{\bullet}_\Lambda(g_1; q_1, q_j)\
\tilde\zeta^{\bullet}_\Lambda(g_2; q_j, q_2).
\end{equation*}
Then, the leading term in \eqref{u2} yields 
\begin{eqnarray}\label{OZ1}
&& \int_{\Lambda^2} dq_1 dq_2\, \phi(q_1) \phi(q_2) \sum_{k=0}^{N-2}P_{N,|\Lambda|}(2+k)\frac{1}{2! k!}
\left[
\sum_{g\in\mathcal B_{2,k+2}}\tilde\zeta^{\bullet}_\Lambda(g;q_1,q_2)+\right.\nonumber\\
&&
\left.
+\sum_{j=3}^{k+2}
\sum_{l=0}^{k-1}
\binom{k-1}{l}\int_{\Lambda} \frac{dq_j}{|\Lambda|}
\sum_{g_1\in \mathcal B_{2,l+2}}\tilde\zeta^{\bullet}_\Lambda(g_1;q_1,q_j)
\sum_{g_2\in \mathcal B^{\text{AF}}_{2,k-l+1}}\tilde\zeta^{\bullet}_\Lambda(g_2;q_j,q_2)
\right].
\end{eqnarray}
We rewrite this in such a way that 
{\it direct two-point correlation function} (uniquely defined up to leading order) as
given in \eqref{c2} appear.
By choosing the label $j=3$ in \eqref{OZ1} we obtain
\begin{eqnarray*}
&& \int_{\Lambda^2} dq_1 dq_2 \,\frac 12 \phi(q_1) \phi(q_2) \left[
\sum_{k=0}^{N-2}P_{N,|\Lambda|}(2+k)\frac{1}{k!}
\sum_{g\in\mathcal B_{2,k+2}}\tilde\zeta^{\bullet}_\Lambda(g;q_1,q_2)+\right.\nonumber\\
&&
\left.
+
\sum_{k=0}^{N-2}P_{N,|\Lambda|}(2+k)
\sum_{l=0}^{k-1}
\int_{\Lambda} dq_3
\frac{1}{l!}
\sum_{g_1\in \mathcal B_{2,l+2}}\tilde\zeta^{\bullet}_\Lambda(g_1;q_1,q_3)
\frac{1}{(k-1-l)!}
\sum_{g_2\in \mathcal B^{\text{AF}}_{2,k-l+1}}\tilde\zeta^{\bullet}_\Lambda(g_2;q_3,q_2)
\right].
\end{eqnarray*}
By using new labels $l_1:=l$ and $l_2:=k-1-l$, 
the last summand can be rewritten as follows
\begin{equation}\label{pOZ}
\sum_{l_1=0}^{N-3}
\int_{\Lambda} dq_3
\frac{1}{l_1!}
\sum_{g_1\in \mathcal B_{2,l_1+2}}\tilde\zeta^{\bullet}_\Lambda(g_1;q_1,q_3)
\sum_{l_2=0}^{N-3-l_1}
P_{N,|\Lambda|}(l_1+l_2+3)
\frac{1}{l_2!}
\sum_{g_2\in \mathcal B^{\text{AF}}_{2,l_2+2}}\tilde\zeta^{\bullet}_\Lambda(g_2;q_3,q_2).
\end{equation}
Let us introduce the following shorthands
\begin{equation}
\bar{C}^{\bullet}_\Lambda(2,l_1+2;q_1,q_3) := 
\frac{1}{l_1!}\hspace{-0.3cm}
\sum_{g_1\in \mathcal B_{2,l_1+2}}\hspace{-0.3cm}\tilde\zeta^{\bullet}_\Lambda(g_1;q_1,q_3)
\end{equation}
and 
\begin{equation} 
\bar{B}^{\bullet}_\Lambda(2,l_2+2;q_3,q_2) := 
\frac{1}{l_2!}\hspace{-0.3cm}\sum_{g_2\in \mathcal B^{\text{AF}}_{2,l_2+2}}\hspace{-0.3cm}\tilde\zeta^{\bullet}_\Lambda(g_2;q_3,q_2).
\end{equation}
Then we can rewrite \eqref{pOZ} as 
\begin{eqnarray}
\int_{\Lambda} dq_3
& & \sum_{l_1=0}^{N-3}  P_{N,|\Lambda|}(l_1+1)\bar{C}^{\bullet}_\Lambda(2,l_1+2;q_1,q_3)
\times \nonumber\\
&&\times  \sum_{l_2=0}^{N-3-l_1}
\frac{P_{N,|\Lambda|}(l_1+l_2+3)}{P_{N,|\Lambda|}(l_1+1)P_{N,|\Lambda|}(l_2+2)}
P_{N,|\Lambda|}(l_2+2) \bar{B}^{\bullet}_\Lambda(2,l_2+2;q_3,q_2),  \label{pOZ2}
\end{eqnarray}
which is a  finite volume version of the convolution term in OZ equation.

\medskip

{\it Proof of Theorem~\ref{thm3}:}
The proof will be divided into two lemmas: the first (Lemma~\ref{lem2.1}) proves the validity of the Ornstein-Zernike equation at
finite volume (up to leading order) and the second (Lemma~\ref{lem2.2}) the infinite volume convergence.
Combining the two results we
conclude the proof of Theorem~\ref{thm3}.
\qed

\medskip

Next we present the two lemmas.
As a consequence of \eqref{pOZ2} we have:
\begin{lemma}\label{lem2.1}
Under the hypothesis of the previous theorems,
the function $c^{(2)}_{\Lambda, N}$ defined in \eqref{c2} fulfils the Ornstein-Zernike equation to leading order in the following sense:
\begin{eqnarray}
\int_{\Lambda^2} \phi(q_1)\phi(q_2) u^{(2)}_{\Lambda, N}(q_1,q_2) \, dq_1  dq_2 & = & \rho^2 
\int_{\Lambda^2} \phi(q_1)\phi(q_2)
c^{(2)}_{\Lambda,N}(q_1,q_2) 
\, dq_1  dq_2 \nonumber\\
&& +  
\int_{\Lambda^2} \phi(q_1)\phi(q_2)
\left(
\int_\Lambda \rho \, c^{(2)}_{\Lambda,N}(q_1,q_3) u^{(2)}_{\Lambda, N}(q_3,q_2) 
dq_3
\right)
\, dq_1  dq_2 \nonumber\\
&& + O\left (\frac{1}{|\Lambda|}\right ). \label{OZfiniteV}
\end{eqnarray}
\end{lemma}

{\it Proof.}
Using the estimates (3.27) and (3.28) in \cite{PT}, 
namely that for some constant $c'$ it holds that for all $l$ and $N$
\begin{equation}
\left| \frac{P_{N,|\Lambda|}(l)}{\rho^{l}} - 1 \right| \leq \frac{c'}{|\Lambda|},
\end{equation}
we can replace in \eqref{pOZ2} all terms of the form $P_{N,|\Lambda|}(l)$ by powers of $\rho$ up to an error of order $O(1/|\Lambda|)$.
Applying that to the fraction $\frac{P_{N,|\Lambda|}(l_1+l_2+3)}{P_{N,|\Lambda|}(l_1+1)P_{N,|\Lambda|}(l_2+2)}$ we replace \eqref{pOZ2} by:
\begin{equation}\label{pOZ3}
\int_{\Lambda} dq_3
\sum_{l_1=0}^{N-3}  P_{N,|\Lambda|}(l_1+1)
\bar{C}^{\bullet}_\Lambda(2,l_1+2;q_1,q_3)  \sum_{l_2=0}^{N-3-l_1}
P_{N,|\Lambda|}(l_2+2) \bar{B}^{\bullet}_\Lambda(2,l_2+2;q_3,q_2).
\end{equation}

We write the Ornstein-Zernike equation plus terms of lower order in $|\Lambda|$.
Then \eqref{pOZ3} can be written as:
\begin{eqnarray*}
&&
 \int_{\Lambda} dq_3
\sum_{l_1=0}^{N-3} P_{N,|\Lambda|}(l_1+1)\bar{C}^{\bullet}_\Lambda(2,l_1+2;q_1,q_3)
\sum_{l_2=0}^{N-3}
P_{N,|\Lambda|}(l_2+2) \bar{B}^{\bullet}_\Lambda(2,l_2+2;q_3,q_2)
\\ &
- &
 \int_{\Lambda} dq_3
\sum_{l_1=0}^{N-3} P_{N,|\Lambda|}(l_1+1)\bar{C}^{\bullet}_\Lambda(2,l_1+2;q_1,q_3)
\sum_{l_2=N-3-l_1}^{N-3}
P_{N,|\Lambda|}(l_2+2) \bar{B}^{\bullet}_\Lambda(2,l_2+2;q_3,q_2).\\
\end{eqnarray*}
We show that the second term is of order $O(1/|\Lambda|)$:
\begin{eqnarray*}
& &
\left| \int_{\Lambda} dq_3
\sum_{l_1=0}^{N-3} P_{N,|\Lambda|}(l_1+1)\bar{C}^{\bullet}_\Lambda(2,l_1+2;q_1,q_3)
\sum_{l_2=N-3-l_1}^{N-3}
P_{N,|\Lambda|}(l_2+2) \bar{B}^{\bullet}_\Lambda(2,l_2+2;q_3,q_2) \right| \\
& \leq & \sup_{q_3'} \sum_{l_2=\lceil N/2 \rceil -2}^{\infty}
P_{N,|\Lambda|}(l_2+2) \left|\bar{B}^{\bullet}_\Lambda(2,l_2+2;q'_3,q_2) \right| \int_{\Lambda} dq_3
\sum_{l_1=0}^{\lfloor N/2 \rfloor } P_{N,|\Lambda|}(l_1+1)\left| \bar{C}^{\bullet}_\Lambda(2,l_1+2;q_1,q_3)\right|
\\
& + & \sup_{q'_3} \sum_{l_2=0}^{\infty}
P_{N,|\Lambda|}(l_2+2) \left|\bar{B}^{\bullet}_\Lambda(2,l_2+2;q'_3,q_2) \right|
 \int_{\Lambda} dq_3
\sum_{l_1=\lceil N/2 \rceil }^{\infty} P_{N,|\Lambda|}(l_1+1)\left| \bar{C}^{\bullet}_\Lambda(2,l_1+2;q_1,q_3)\right|.
\end{eqnarray*}
In order to show that the above bound is of order  $O(1/|\Lambda|)$,
one notes that both summands contain the following two factors which are tails of the corresponding convergent series:
\begin{equation}\label{hyp1}
\sup_{q_2,q_3} \sum_{l=N+1}^{\infty} P_{N,|\Lambda|}(l+2) \left|\bar{B}^{\bullet}_\Lambda(2,l+2;q_3,q_2) \right| \leq Ce^{-cn}
\end{equation}
and
\begin{equation}\label{hyp2}
\sup_{q_1} \sum_{l=N+1}^{\infty} P_{N,|\Lambda|}(l+1)
\int_{\Lambda} dq_3
 \left| \bar{C}^{\bullet}_\Lambda(2,l+2;q_1,q_3) \right| \leq C e^{-cn},
\end{equation}
for some constants $C,c>0$.
The first follows from the bound in \eqref{boundforBbullet}, while the second is 
claimed in \eqref{boundforCbullet} and proved in the next lemma.
\qed

The second result is about the convergence and integrability of $c^{(2)}_{N}(q_1,q_2)$ as $N\to\infty$.
In order to take the limit in (\ref{OZfiniteV}) and get the infinite volume
version of the OZ equation, we need to prove \eqref{boundforCbullet}
which is given in the following lemma:

\begin{lemma}\label{lem2.2}
For some positive constants $C$ and $c$ independent of $N$ and $\Lambda$ and 
for every $l_1\in \mathbb N$ and $q_1 \in \Lambda$ we have that
\begin{equation}
P_{N,|\Lambda|}(l_{1}+1) \int_{\Lambda} dq_2 
\left| \bar{C}^{\bullet}_\Lambda(2,l_1+2;q_1,q_{2}) \right|  \leq C \rho e^{-c l_{1}},
\end{equation}
for $\Lambda$ large enough.
\end{lemma}

\begin{remark}
As it will be clear in the proof, for the above estimate to hold it is important that we have an integral in $q_2$, that is an integral over the variable corresponding to the second white vertex. 
For short we call it the \emph{integrated white vertex}.
\end{remark}

{\it Proof.}
The proof follows the line of calculation in Lemma~\ref{exchange}. 
The main difference is that here we 
do not require that there exists a special polymer $V_0$ containing both white vertices. 
Hence we restrict to the class
\begin{eqnarray}
&& I(V)=1,\,\forall V\in\supp I,\,\,\, \text{and}\label{p3.3}\\
&& m+k= \sum_{V\in\supp I}(|V|-1)+1\label{p3.4}
\end{eqnarray}
and we denote it by using the superscript $**$ over the sum, in order to distinguish it from the previous case.
Recalling the shortcut $I\sim g$ for the class 
of multi-indices in $\mathcal I(\mathcal V_{2,l_1})$ as in \eqref{beforetheend},
we have:
\begin{eqnarray}\label{secondeq}
&& P_{N,|\Lambda|}(l_{1}+1) \int_{\Lambda} dq_2 
\left| \bar{C}^{\bullet}_\Lambda(2,l_1+2;q_1,q_{2}) \right| =\nonumber\\
& = &
\frac{N(N-1)\ldots (N-(l_1+1)+1)}
{|\Lambda|^{l_{1}+1}}
\int_{\Lambda} dq_2 \left| \frac{1}{l_1!}\sum_{g\in\mathcal C_{2,2+l_{1}}}
\tilde\zeta_\Lambda^{\bullet}(g;q_{1},q_{2}) \sum^{**}_{I\sim g} c_I \right| \nonumber \\
 &=&
\frac{N}{|\Lambda|}
\binom{N-1}{l_{1}} \int_{\Lambda} dq_2 
\left| \sum^{**}_{\substack{I\in \mathcal I(\mathcal V_{2,l_1})\\ A(I) = [l_1+2]}} c_{I} \frac{1}{|\Lambda|^{l_{1}}}\sum_{\substack{g\in\mathcal C_{2,2+l_{1}}: \\ g \sim I}}
\tilde\zeta_\Lambda^{\bullet}(g;q_{1},q_{2}) \right| .
\end{eqnarray}
The class $\mathcal V_{2,l_{1}}$ consists of all
subsets of the labels corresponding to 
the white vertices $\{1, 2\}$ and the black vertices $\{3, \ldots , l_1+2\}$.
The class $g\sim I$ is as before in \eqref{eqclustq}.
The compatibility graph of the polymers 
is a connected graph whose blocks are complete graphs (usually called Husimi graphs, see \cite{L, H}). 
Within this structure we denote by $V_1, \ldots, V_r$ the chain of pairwise incompatible 
polymers such that the label 
$1 \in V_1$  and the label $2 \in V_r$. Note that $r$ could be equal to $1$, but in this case the structure would be exactly as in the previous theorem.
We denote by $i_j$ the common label of $V_j$ and $V_{j+1}$, $j=1,\ldots, r-1$ 
and by $V'_s$, for $s$ from an index set $S$, the 
remaining polymers attached to the rest of the structure
by the label $i_{s}$. 
Note that by translation invariance the activity associated to $V'_s$ does not depend on the label that 
connects it to the chain. 
Hence we can write (letting $x_{i_{0}}:=q_{1}$ and $x_{i_{r}}:=q_{2}$)
\begin{equation}
\sum_{\substack{g\in \mathcal C_{2,2+l_1}: \\ g\sim I}}
 \tilde\zeta_\Lambda^{\bullet}(g;q_1,q_{2})
 =  \int_{\La^{r-1}} \prod_{j=1}^{r-1}dx_{i_j} \prod_{j=1}^{r }\sum_{g\in\mathcal C_{V_{j}}}\tilde\zeta_\Lambda^{\bullet}(g;x_{i_{j-1}},x_{i_{j}}) \prod_{s \in S} \sum_{g\in\mathcal C_{V_{s}}}\tilde\zeta_\Lambda^{\bullet}(g;x_{i_s}) .
\end{equation}

Notice that this expression does not factorise like in the previous case for the 
reason that the two white vertices are not in the same polymer.
It is exactly here that the extra integral over $dq_{2}$ is helpful.
By integrating over the common labels $i_j$, $j=1,\ldots, r-1$, we obtain:
\begin{eqnarray*}
& &
\int_\Lambda dq_2 \left| \sum_{\substack{g\in \mathcal C_{2,2+l_1}: \\ g\sim I}}
 \tilde\zeta_\Lambda^{\bullet}(g;q_1,q_{2})
\right|\\
 & \leq &
\int_\Lambda dq_2  \int_{\La^{r-1}} \prod_{j=1}^{r-1} dx_{i_j} \prod_{j=1}^{r }|\sum_{g\in\mathcal C_{V_{j}}}\tilde\zeta_\Lambda^{\bullet}(g;x_{i_{j-1}},x_{i_{j}})| \prod_{s \in S} |\sum_{g\in\mathcal C_{V_{s}}}\tilde\zeta_\Lambda^{\bullet}(g;x_{i_s})| .\end{eqnarray*}
Transforming to the difference variables $x_{i_{j}}-x_{i_{j-1}}$ we see that the integrals in the chain factorize as well. 
Then, by introducing the notation 
\begin{equation}\label{zetahat2}
\bar\zeta_{\Lambda}^{\bullet}(V) :=
  \left\{ \begin{array}{ll} 
  \frac{1}{|\Lambda|^{|V|-2}}\sup_{q_1 \in \Lambda} \int_\Lambda dq_2|\sum_{g\in\mathcal C_{V}}
 \tilde\zeta_{\Lambda}^{\bullet}(g;q_1,q_2)| ,
  & \mbox{if } V \supset \{q_1,q_2\}, \\
  \frac{1}{|\Lambda|^{|V|-1}}\int_\Lambda dq_{2}|\sum_{g\in\mathcal C_{V}}
 \tilde\zeta_{\Lambda}^{\bullet}(g;q_2)|,
  & \mbox{if } V \ni q_2, V\cap\{q_{1}\}=\varnothing\\
\frac{1}{|\Lambda|^{|V|-1}}\sup_{q_1 \in \Lambda} | \sum_{g\in\mathcal C_{V}}
 \tilde\zeta_{\Lambda}^{\bullet}(g;q_1) |, & \mbox{if } V \ni q_1, V\cap\{q_{2}\}=\varnothing\\
\frac{1}{|\Lambda|^{|V|}} | \sum_{g\in\mathcal C_{V}}
 \tilde\zeta_{\Lambda}^{\bullet}(g;\varnothing) |, & \mbox{if } V \cap \{ q_1, q_2 \} = \varnothing ,
  \end{array} \right.
\end{equation}
we obtain the following upper bound for \eqref{secondeq}:
\begin{equation}\label{froml1toN}
\rho\sum_{\substack{I\in\mathcal I(\mathcal V_{2,l_1})\\A(I) = [l_1+2]}}^{*}|c_{I}|\binom{N-1}{l_{1}}\prod_{V\in\supp I}|\bar\zeta_{\Lambda}^{\bullet}(V)|=
\rho
\sum_{\substack{A\subset [N-1]\\ |A|=l_1}}
\sum_{I: A(I)=A\cup\{1,2\}}
^{*} |c_I|
\prod_{V\in\supp I}|\bar\zeta_{\Lambda}^{\bullet}(V)|.
\end{equation}
Then, it is easy to show that the abstract polymer model in $\mathcal V_{2,N-2}$
(with $2$ white labels and $N-2$ black)
and activities $\bar\zeta_\Lambda^\bullet$
satisfies the hypothesis \eqref{7.4} of Theorem~\ref{thmCE} (by obtaining similar bounds as previously). 
Thus, from \eqref{froml1toN}, using \eqref{7.5},
we obtain that
\be
\sum^*_{\substack{I\in\mathcal I(\mathcal V_{2,N-2}) \\ |A(I)\setminus\{1,2\}|=l_1}}
|c_I| |\bar\zeta^\bullet_\Lambda|^I
\leq e^{-c l_{1}}
\sum^*_{\substack{I\in\mathcal I(\mathcal V_{2,N-2}) \\  |A(I)\setminus\{1,2\}|=l_1}}
|c_I|  |\bar\zeta^\bullet_\Lambda|^I e^{c l_{1}} \leq C e^{-c l_{1}},
\ee
for some $C>0$ as in \eqref{estforF}.
\qed

\section{Towards a combinatorial interpretation}\label{comb}

Until recently, it was customary to investigate
the density expansions of thermodynamic quantities in the
context of the grand-canonical ensemble. This was because the lack
of the canonical constraint (i.e., having a fixed number of particles) allowed
for special re-summations.
As a result, the representation of coefficients is given by classes of graphs whose different connectivity properties are related to combinatorial identities, see \cite{S} for more details.
For example, for the conjugate pair of free energy and pressure this is the well known dissymmetry theorem, see e.g. Theorem~3.7 in \cite{L}. 
The correlation functions $h^{(n)}$ for $n\geq 2$ actually correspond to an easier structure than in the case $n=1$.
Let us consider the case $n=2$ and the expansion of $\rho^{(2)}$ in terms of the activity. Upon the removal of one white vertex the graph decomposes into connected components which either contain the other white vertex or not. Collecting those not containing a white vertex, we reconstruct the expansion of $\rho^{(1)}$ in terms of the activity. 
One repeats the same procedure for the other white vertex. The remaining graph has the property that both white vertices are contained in exactly one articulation point free block. Considering the associated block-articulation point graph, the parts that do not correspond to the special block containing the white vertices, reconstruct exactly the  
$\rho^{(1)}$-expansion at each black vertex of the special block. One can argue similarly for all $n \geq 2$, cf. Section~5 in \cite{S}. Using the combinatorial language, as e.g. in \cite{L}, this is just the following combinatorial identity interpreted as formal power series:
\begin{equation}
\mathcal{C}^{*}_n = \left(\mathcal{C}^\bullet \right)^n\mathcal{B}^{\text{AF}}_n(\mathcal{C}^\bullet),
\end{equation}
where $\mathcal{C}^{*}_n$, $\mathcal{B}^{\text{AF}}_n$ respectively, denotes the set of connected, articulation point free respectively, graphs with $n$ special vertices. $\mathcal{C}^\bullet$ denotes the set of graphs with
one special vertex, but multiplied with the activity.

The case $n=1$ has a more difficult structure. Let us derive it in more detail; we have
\begin{eqnarray}\label{gconebody}
\frac{\rho^{(1)}_{\Lambda}(q_1)}{z} & = & \frac{1}{\Xi_{\Lambda}(z)}
\sum_{n\geq 1}\frac{z^{n-1}}{(n-1)!}\int_{\Lambda^{n-1}}dq_2\ldots dq_n e^{-\beta H_{\Lambda}({\bf q})},
\end{eqnarray}
where $z$ is the activity and $\Xi_{\Lambda}(z)$ the grand-canonical
partition function.
Writing $^{-\beta H_{\Lambda}({\bf q})} = \sum_{g \in \mathcal{G}_n} \prod_{\{i,j\} \in E(g)} f_{i,j}$, we split the graph and the integral over the connected components of each graph. Recalling the definition of the activity
$\tilde\zeta^{\bullet}_{\Lambda}(g;q_1,\ldots , q_n)$ given in \eqref{act_with_points}, 
we get that   \eqref{gconebody} equals to
\begin{eqnarray}\label{gconebody2}
&& \frac{1}{\Xi_{\Lambda}(z)}
\sum_{n\geq 1}\frac{1}{(n-1)!} \sum_{k\geq 1}\frac{1}{k!}\sum_{(P_0, \ldots P_k)\in\Pi(2,\ldots,n)} \left(z^{|P_0|} \!\!\!\!\!\sum_{g \in \mathcal{C}(P_{0}\cup \{1\})} \!\!\!\!\!\! \tilde\zeta^{\bullet}_{\Lambda}(g;q_1) \right) \prod_{j=1}^k \left(z^{|P_j|}\!\!\!\!\! \sum_{g \in \mathcal{C}(P_{j})}\!\!\!\! \tilde\zeta^{\bullet}_{\Lambda}(g;\varnothing) \right) \nonumber
\\
& = & \sum_{n\geq 1} n z^{n-1}
\frac{1}{n!}\sum_{g\in\mathcal C_n}\tilde\zeta^{\bullet}_{\Lambda}(g;q_1)
 =  \sum_{n\geq 1} \frac{z^{n-1}}{n!}\sum_{g\in\mathcal C_{1,n}}
 \tilde\zeta^{\bullet}_{\Lambda}(g;q_1), 
\end{eqnarray}  
where $1$ is a special point (hence absorbing the factor $n$).
Upon the removal of the white vertex, the remaining graph splits into 
connected components $P_1, \ldots , P_k$.
Denote by $\mathcal{C}_{1}(P)$ the set of all graphs in $P \cup \{ 1\}$ which have
$1$ as a special vertex and are still connected even on its removal. In other words, 
$1$ is not an articulation circle in the sense of Stell.
Then, from \eqref{gconebody}, 
we have that
\begin{eqnarray}\label{gconebody3}
\frac{\rho^{(1)}_{\Lambda}(q_1)}{z} 
& = & 1+ \sum_{n=1}^\infty \frac{1}{n!} \sum_{k=1}^n \frac{1}{k!}\sum_{(P_1, \ldots P_k)\in\Pi(1,\ldots,n)} \prod_{j=1}^k \left(z^{|P_j|} \sum_{g \in \mathcal{C}_{1}(P_{j})}
\tilde\zeta^{\bullet}_{\Lambda}(g;q_1)\right)\nonumber\\
& = & 1+ \sum_{k\geq 1}\frac{1}{k!}\Big(
 \sum_{p\geq 1} \frac{z^{p}}{p!} \sum_{g \in \mathcal{C}_{1,p+1}}
 \tilde\zeta^{\bullet}_{\Lambda}(g;q_1)
 \Big)^{k}
 \nonumber\\
& = &  \exp\left\{ \sum_{g \in \mathcal{C}^{*}_{1}}
\frac{z^{|g|}}{|g|!}
\tilde\zeta^{\bullet}_{\Lambda}(g;q_1)
\right\},
\end{eqnarray}
where in the last sum we denote by $\mathcal{C}^*_{1}$ the set of connected graphs with $1$ as a 
special vertex and any cardinality.
At this point, as described in the case $n\geq 2$ before, we are able to systematically replace the black $z$ vertices by black $\rho^{(1)}_{\Lambda}$ vertices and thus obtain that
\begin{equation}
\sum_{g \in \mathcal{C}^*_{1}}
\frac{z^{|g|}}{|g|!}
\tilde\zeta^{\bullet}_{\Lambda}(g;q_1) =  \sum_{m\geq 1}\beta_{\Lambda, m}(\rho^{(1)}_{\Lambda}(q_{1}))^{m}= F_{\Lambda}'(\rho^{(1)}_{\Lambda}(q_1)),
\end{equation}
recalling that 
\be\label{n4}
\beta_{\Lambda, m}:=
\frac{1}{m!}
\sum_{g\in\BB_{1,m+1}}\int_{\Lambda^{m}}\prod_{\{i,j\}\in E(g)}(e^{-\beta V(q_i-q_j)}-1)
dq_2\ldots dq_{m+1},\quad q_1 \text{ fixed},
\ee
is the virial coefficient and $F_\Lambda(\rho):=\sum_{m\geq 1}\frac{1}{m+1}\beta_{\Lambda,m}\rho^{m+1}$.

This is exactly the combinatorial identity given in \cite{L}, Theorem~$1.1$. 
The above calculation is also one of the motivations to define (following \cite{S}): \begin{equation}\label{h1}
h^{(1)}(q_1):=\log(\rho(q_1)) - \log(z)
= \sum_{m\geq 1}\beta_{m}(\rho(q_{1}))^m ,
\end{equation}
in the thermodynamic limit.
Note that because of translation invariance both $h^{(1)}(q_1)$ and $\rho(q_1)$ are constant.
This is also closely related to the Legendre transform giving the equivalence of ensembles between pressure and free energy at the thermodynamic limit:
\begin{equation*}
p(z)=\sup_{\rho}\{\rho\log z-f(\rho)\},\qquad
f(\rho)=\sup_z\{\rho\log z - p(z)\}.
\end{equation*}
In the first case the $\sup$ is attained at $\log z=f'(\rho)$ and hence 
\begin{equation}
h^{(1)} = \log \rho - f'(\rho) = F'(\rho),
\end{equation}
where $F(\rho) = \rho (\log \rho -1) -f(\rho)$ is the free energy corresponding to the ``interaction"
between the particles.

\medskip

We conclude this section by noting that
the OZ equation corresponds to the following easy combinatorial fact. For the second correlation functions the expansion in the density is given by the sum over all graphs free of articulation vertices. Hence the block graph associated to such a graph is  actually a chain connecting the two white vertices. The OZ equation is nothing more than an iterative representation of this fact.

\section{Applications to liquid state theory}\label{LST}
The rigorous expansions that we present in this paper can serve as a tool for quantifying the error in
existing theories which are extensively used in the liquid state,
as well as for suggesting systematic error-improving schemes.
We give here a first glimpse of this. To start, we recall that
the Ornstein-Zernike equation \eqref{OZv1} is not a closed equation as it involves both
correlation functions $h^{(2)}(q_1,q_2)$ and $c(q_1,q_2)$.
One suggestion for a closure is the Percus-Yevick (PY) equation \cite{PY} that we describe 
below.
Starting from the OZ equation for $h^{(2)}(r)$ and $c(r)$,
following \cite{SPY}, one first introduces a new function $t$ as follows:
\begin{equation}\label{OZ}
 t(r) : =c*h^{(2)}(r),
\end{equation}
where we use the convolution: $c*h^{(2)}(r) := \rho \int c(r')h^{(2)}(r-r') dr'$.
Then the OZ equation takes the form
\begin{equation}\label{OZ4}
h^{(2)}(r)=c(r)+t(r).
\end{equation}
Note that all involved functions ($h^{(2)}$, $c$ and $t$) are analytic functions in $\rho$.
Furthermore, $c(r)$ can be written as
\begin{equation}\label{cofr}
c(r)=f(r)(1+t(r))+m(r),
\end{equation}
where $f(r):=e^{-\beta V(r)}-1$ is a known function of the potential $V(r)$.
The relation \eqref{cofr} is essentially the definition of $m(r)$ which is an analytic function of $\rho$ as well. Following \cite{SPY} the function $m$  
can be expressed as a sum over two connected graphs which upon removal of the direct link $f$ connecting
the white vertices (if it is present) it is two-connected (no articulation and no nodal points). 
For example, the first term of $m(r)$ is the graph $1-3-2-4-1$.
However, in \cite{SPY}, {\it``the manipulations involved in obtaining these
infinite sums ... have been carried out in a purely formal way and we have not examined 
the important but difficult questions of convergence and the legitimacy of the
rearrangement of terms"}.
The present paper establishes this convergence
with respect to $f$-bonds.
The convergence allows to quantify the error after truncating these terms. For example, $m$ is of order $\rho^2$.
Furthermore, a future plan is to investigate whether
another suggestion could be made, relating some of the terms
in $m(r)$ with respect to $t(r)$, or by introducing another function (instead of $t(r)$)
as a candidate for a good choice for ``closing" OZ equation.
Combining \eqref{OZ} with \eqref{OZ4} and \eqref{cofr} we obtain:
\begin{equation}\label{combined}
t=[f(1+t)+m]*[f(1+t)+m]
+[f(1+t)+m]*t.
\end{equation}
One version of PY equation is setting $m(r)\equiv 0$ and obtaining
a closed equation for $t(r)$.

Alternatively, using \eqref{OZ4} and \eqref{cofr}  one can introduce the functions $y(r)$ and $d(r)$ by
\begin{equation}\label{yandd}
g^{(2)}(r) = e^{-\beta V(r)} (1+t(r)) +m(r)=:e^{-\beta V(r)} y(r) ,\qquad y(r)=:1+t(r)+d(r),
\end{equation}
and hence $m(r)=e^{-\beta V} d(r)$. Thus,
we can rewrite \eqref{combined} as
\begin{eqnarray}\label{combined2}
y & = & 1+ d + [f\ y+d]*[e^{-\beta V}y -1].
\end{eqnarray}
Again, setting $d(r)\equiv 0$ we obtain another version of PY equation. All
involved functions are analytic in $\rho$ and our results imply that the formal order in $\rho$ of $d$ coincides with the actual order.
Now, one can investigate a method of systematically improving the PY equation, by adding 
some terms from $d$ (or 
from $m$ for hard-core potentials).
For example, in \cite{SPY}  it was suggested to set $d$ equal to the first order term in its expansion, 
since this gives
a {\it ``PY approximation that it leads to an approximate $g$ that is exact
through terms of order $\rho^2$ in its virial expansion"}.
 A partial goal of the analysis in the present paper is to
provide a framework in which one can further investigate
such closure schemes and estimate the relevant error.

Other closures include the {\it Hypernetted Chain} (HNC) equation, the Born-Green-Yvon (BGY) hierarchy
and many others for which we could investigate the validity of the corresponding
graphical expansions.
We conclude by mentioning that another direction that has attracted considerable interest 
is the construction of exact solutions
of the PY equation, which however usually cannot be expressed as truncations of
convergent series. But still, several suggestions have been made for models of rigid spheres;
see \cite{CS} and the references therein for a comparison of the different methods.

\bigskip

{\bf Acknowledgments.}
We acknowledge support from the London Mathematical Society via a research in pairs Scheme 4 grant.
T. K. acknowledges support from J. Lebowitz via
NSF Grant DMR 1104501 and AFOSR Grant F49620-01-0154.

\end{document}